\documentclass[aps, prd, twocolumn, preprintnumbers, superscriptaddress, nofootinbib]{revtex4-2}

\usepackage{style}

\begin{document}

\title{Dark Matter Direct Detection in Materials with Spin-Orbit Coupling}

\author{Hsiao-Yi Chen}
\affiliation{Department of Applied Physics and Materials Science, California Institute of Technology, Pasadena, CA 91125, USA}
\affiliation{RIKEN Center for Emergent Matter Science (CEMS), Wako, Saitama, 351-0198, Japan}
\author{Andrea Mitridate}
\affiliation{Walter Burke Institute for Theoretical Physics, California Institute of Technology, Pasadena, CA 91125, USA}
\author{Tanner Trickle}
\affiliation{Walter Burke Institute for Theoretical Physics, California Institute of Technology, Pasadena, CA 91125, USA}
\author{Zhengkang Zhang}
\affiliation{Department of Physics, University of California, Santa Barbara, CA 93106, USA}
\author{Marco Bernardi}
\affiliation{Department of Applied Physics and Materials Science, California Institute of Technology, Pasadena, CA 91125, USA}
\author{Kathryn M. Zurek}
\affiliation{Walter Burke Institute for Theoretical Physics, California Institute of Technology, Pasadena, CA 91125, USA}

\preprint{CALT-TH-2022-007}

\begin{abstract} 
Semiconductors with $\mathcal{O}(\text{meV})$ band gaps have been shown to be promising targets to search for sub-MeV mass dark matter (DM). In this paper we focus on a class of materials where such narrow band gaps arise naturally as a consequence of spin-orbit coupling (SOC). Specifically, we are interested in computing DM-electron scattering and absorption rates in these materials using state-of-the-art density functional theory (DFT) techniques. To do this, we extend the DM interaction rate calculation to include SOC effects which necessitates a generalization to spin-dependent wave functions. We apply our new formalism to calculate limits for several DM benchmark models using an example ZrTe$_{5}$ target and show that the inclusion of SOC can substantially alter projected constraints. 
\end{abstract}

\maketitle
\tableofcontents

\section{Introduction}
\label{sec:introduction}

Detection of dark matter (DM) through non-gravitational interactions remains one of the main goals of particle physics. Electronic excitations have been identified as a promising path to lead the direct detection of DM to sub-GeV masses, a region not kinematically accessible in experiments based on nuclear recoil. A variety of avenues to search for DM induced electronic excitations have been proposed: ionization in noble gases~\cite{Essig:2011nj,Graham:2012su,Lee:2015qva,Essig:2017kqs,Catena:2019gfa,Agnes:2018oej,Aprile:2019xxb,Aprile:2020tmw}, excitations across a band gap in crystal targets~\cite{Essig:2011nj,Graham:2012su,Essig:2012yx,Lee:2015qva,Essig:2015cda,Derenzo:2016fse,Hochberg:2016sqx,Bloch:2016sjj,Kurinsky:2019pgb,Trickle:2019nya, Griffin:2019mvc,Griffin:2020lgd,Du:2020ldo, Mitridate:2021ctr}, superconductors~\cite{Hochberg:2015pha, Hochberg:2016ajh}, graphene \cite{Hochberg:2016ntt}, Dirac materials~\cite{Hochberg:2015fth, Hochberg:2017wce, Coskuner:2019odd,Geilhufe:2019ndy,Inzani:2020szg}, and transitions between molecular orbitals in aromatic organics~\cite{Blanco:2019lrf, Blanco:2021hlm}. 

In this work we focus on a specific class of semiconductors for which $\mathcal{O}(\si{\meV})$ band gaps (as opposed to typical $\mathcal{O}(\si{\eV})$ band gaps in semiconductors and insulators) arise as a consequence of spin-orbit coupling (SOC) effects. Targets with such small band gaps can probe DM masses down to $\mathcal{O}(\si{\keV})$ via scattering, and $\mathcal{O}(\si{\meV})$ via absorption, while still suppressing thermal noise. Moreover, some of these SOC materials have tunable band structures, a property which makes them interesting candidates for direct detection experiments~\cite{Inzani:2020szg}. 

However, SOC effects introduce some intricacies in the DM-electron interaction rate calculations since the Bloch wave functions are no longer eigenstates of the $S_z$ operator, and therefore become two-component objects in spin space. This implies that electron spin sums cannot be trivially performed, and new transition form factors must be computed. For example, spin-dependent vector mediated scattering can no longer be related to its spin-independent counterpart, and must be computed from first principles. We extend the framework in Sec.~\ref{sec:dm_interaction_rate_formalism} and implement the new spin-dependent form factors numerically within \href{https://exceed-dm.caltech.edu}{\textsf{EXCEED-DM}}~\cite{EDC2021, Griffin:2021znd}, which is publicly available on Github \href{https://github.com/tanner-trickle/EXCEED-DM}{\faGithub}.

To showcase the formalism developed in this paper, we apply it to a target with important SOC effects, \ce{ZrTe5}. This material has been extensively studied in the context of DM direct detection~\cite{Coskuner:2019odd, Geilhufe:2019ndy, Hochberg:2017wce} as a leading candidate for Dirac material targets. Dirac materials are characterized by low-energy excitations which behave like free electrons and satisfy the Dirac equation. The properties of the electronic excitations can then be understood by a simple extension of the standard QED results. An additional consequence is that they have weak electromagnetic screening, even with a small energy gap between the valence and conduction bands, making them a desirable target for sub-MeV dark matter coupled to electrons via a dark photon mediator \cite{Coskuner:2019odd}. In this work, however, we will focus only on ZrTe$_{5}$ properties which stem from its SOC nature, and we will not exploit any of the ones deriving from its Dirac nature (which is still debated \cite{Zheng2016, Chen2015, Liu2018, Li:2014bha, Chen2015a, Yuan2016, Chen2017, Monserrat2019, Wu2016, Nair2017, Zhang2017, Moreschini2016}). 

To illustrate the variety of DM models that an SOC target can probe, we consider several different DM models and processes. Specifically, we will study:
\begin{itemize}
	\item Standard spin-independent (SI) and spin-dependent (SD) scattering via vector mediators. The fundamental interaction Lagrangians for these models take the form
	\begin{align}\label{eq:scatt_lagr}
	\def\arraystretch{1.5}
		\mathcal{L}_{\rm int}=\left\{\begin{array}{ll}
		\phi_\mu\left(g_\chi \overline\chi\gamma^\mu\chi+g_e\overline\psi\gamma^\mu\psi\right) & \quad\rm (SI)\\
		\phi_\mu\left(g_\chi \overline\chi\gamma^\mu\gamma^5\chi+g_e\overline\psi\gamma^\mu\gamma^5\psi\right) & \quad\rm (SD)
		\end{array}\right.
	\end{align}
	where $\psi$ and $\chi$ are the electron and DM fermion fields, respectively, and $\phi_\mu$ is the dark mediator field. 
	\item Scalar, pseudoscalar, and vector DM absorption. In this case the fundamental interaction Lagrangians take the form
	\begin{align}\label{eq:abs_lagr}
		\def\arraystretch{1.5}
		\mathcal{L}_{\rm int}=\left\{\begin{array}{ll}
		g_e \phi\overline\psi \psi & \quad\rm (scalar\;DM)\\
		g_e \phi\overline\psi i \gamma^5\psi &\quad\rm (pseudoscalar\; DM)\\
		g_e \phi_\mu \overline\psi\gamma^\mu\psi & \quad\rm (vector\;DM)\,.
		\end{array}\right.
	\end{align}
\end{itemize} 

The paper is organized as follows. In Sec.~\ref{sec:dm_interaction_rate_formalism} we generalize the DM interaction rate formalism to account for spin-dependent wave functions in general (spin-orbit coupled, anisotropic) targets. Then in Sec.~\ref{sec:ZrTe5_results} we apply these results to the candidate material \ce{ZrTe5} and compare the results obtained with and without the inclusion of SOC effects. 
Further details of DFT calculation are presented in App.~\ref{subapp:numerics_dft}, and convergence tests for the results shown in Sec.~\ref{sec:ZrTe5_results} can be found in App.~\ref{subapp:numerics_dm_int_rate_convergence}.

\section{DM Interaction Rate Formalism}
\label{sec:dm_interaction_rate_formalism}

In this section we derive the rates for transitions between electronic energy levels induced by DM absorption and scattering. For the targets of interest here, the electronic energy levels can be labelled by a band index $i$ and a momentum $\vec{k}$ within the first Brillouin zone (1BZ), which we collectively indicate with an index $I=\{i,\vec{k}\}$. The wave functions of the electronic states can be written in the Bloch form as:
\begin{align}
	\Psi_{I}(\vec{x})=\frac{1}{\sqrt{V}}\, e^{i\vec{k}\cdot\vec{x}}\, \mathbf{u}_I(\mathbf{x})
    \label{eq:bloch_wf}
\end{align}
where the periodic Bloch wave functions $\mathbf{u}_I$ are two-component vectors in the spin basis, and $V$ is the crystal volume. 

\subsection{Absorption}
\label{subsec:formalism_absorption}

In this subsection, we use the non-relativistic (NR) effective filed theory (EFT) developed in Ref.~\cite{Mitridate:2021ctr}, and summarized in Appendix \ref{app:self_ene}, to compute DM absorption rates in materials with sizable SOC. 

The absorption rate of a state can be derived from the imaginary part of its self-energy. In a medium, care must be taken due to the possible mixing between the DM, $\phi$, and SM states (in our case the SM photon, $A$). In the presence of such mixing effects, the DM absorption rate is related to the imaginary part of the self-energy of the ``mostly DM" eigenstate, $\Pi_{\hat\phi\hat\phi}$:
\begin{align}\label{eq:abs_rate}
	\Gamma_{\rm abs}^\phi=-\frac{Z_{\hat\phi}}{\omega}\,{\rm Im}\,\Pi_{\hat\phi\hat\phi} \,,
\end{align}
where $\omega\simeq m_\phi$ is the energy of the DM state, and $Z_{\hat\phi}=\bigl(1-\frac{d\,{\rm Re}\Pi_{\hat\phi\hat\phi}}{d\omega^2}\bigr)^{-1}=1+\mathcal{O}(g_e^2)$ is the wave function renormalization which we will approximate as unity in the following. The total absorption rate per unit target mass, $R$, is given by 
\begin{align}\label{eq:tot_rate}
	R=\frac{\rho_\phi}{\rho_T m_\phi} \frac{1}{n} \sum_{\eta=1}^{n} \Gamma_{\rm abs}^{\phi_\eta}
\end{align}
where $n$ is the number of degrees of freedom of the DM particle ($n = 3$ for vector DM and $n = 1$ for scalar and pseudoscalar DM) and we average over the incoming DM polarizations. The DM density, $\rho_\phi$, is taken to be $0.4\,\mathrm{GeV}\,\mathrm{cm}^{-3}$, and $\rho_T$ is the target density. 

To derive $\Pi_{\hat\phi\hat\phi}$ we need to diagonalize the in-medium self-energy matrix, which in our case contains a mixing between the DM and the SM photon:
\begin{widetext}
    \begin{align}
        \label{eq:self_energy_matrix}
        \mathcal{S}_\text{eff} \supset -\frac{1}{2} \int d^4Q \,
        \begin{pmatrix} A^\lambda & \phi^\eta \end{pmatrix}
        \begin{pmatrix} \Pi_{AA}^{\lambda\lambda'} & \Pi_{A\phi}^{\lambda\eta'} \\ \Pi_{\phi A}^{\eta\lambda'} & m_\phi^2\,\delta^{\eta\eta'} + \Pi_{\phi\phi}^{\eta\eta'} \end{pmatrix}
        \begin{pmatrix} A^{\lambda'} \\ \phi^{\eta'} \end{pmatrix}\,,
    \end{align}
\end{widetext}
where the implicit sum over $\lambda,\lambda'$ ($\eta,\eta'$) runs over the photon (DM) polarizations, and we have introduced the self-energies polarization components defined as:
\begin{align}
\Pi^{\lambda\lambda'}\equiv\vec{\epsilon}_\mu^\lambda \Pi^{\mu\nu}\vec{\epsilon}_\nu^{\lambda'*}
\end{align}
where $\vec{\epsilon}_\mu^\lambda$ are polarization vectors. In general, the polarization vectors which diagonalize this matrix are not the typical longitudinal and transverse polarization vectors, since mixing can occur (i.e., $\Pi_{AA}^{L,T}\ne0$). However one can always find an appropriate basis to diagonalize the DM and photon self-energies. In this basis Eq.~\eqref{eq:self_energy_matrix} becomes
\begin{widetext}
    \begin{align}
        \label{eq:diag_self_en}
        \mathcal{S}_\text{eff} \supset -\frac{1}{2} \int d^4Q 
        \begin{pmatrix} A^\lambda & \phi^\eta \end{pmatrix}
        \begin{pmatrix} \Pi_{AA}^{\lambda}\,\delta^{\lambda\lambda'} & \Pi_{A\phi}^{\lambda\eta'} \\ \Pi_{\phi A}^{\eta\lambda'} & (m_\phi^2 + \Pi_{\phi\phi}^\eta)\,\delta^{\eta\eta'} \end{pmatrix}
        \begin{pmatrix} A^{\lambda'} \\ \phi^{\eta'} \end{pmatrix}\,,
    \end{align}
\end{widetext}
where $\Pi_{AA}^{\lambda}$ and $\Pi_{\phi\phi}^{\eta}$ are the eigenvalues of $\Pi_{AA}^{\lambda\lambda'}$ and $\Pi_{\phi\phi}^{\eta\eta'}$ respectively. 

The off-diagonal terms in Eq.~\eqref{eq:diag_self_en} are perturbatively suppressed by a factor of $g_e$ with respect to the $\Pi_{AA}$ terms. Therefore, working at order $\mathcal{O}(g_e^2)$, we find that the in-medium self-energy for the $\eta$ polarization of the mostly DM eigenstate is given by:
\begin{align}
	\label{eq:diag_se}
	\Pi_{\hat\phi\hat\phi}^\eta= \Pi_{\phi\phi}^\eta+\sum_\lambda\frac{\Pi_{\phi A}^{\eta\lambda}\Pi_{A\phi}^{\lambda\eta}}{m_\phi^2-\Pi_{AA}^\lambda}\,.                           
\end{align}
Since vector DM couples to electrons in the same way as the photon, one can derive the relevant self-energies by simply replacing the electromagnetic charge with $g_e$, e.g., $\Pi_{\phi A}^{\eta\lambda} = -(g_e/e) \,\Pi_{AA}^{\lambda} \,\delta^{\eta \lambda}$. Doing so allows us to write Eq.~\eqref{eq:diag_se} in terms of the photon self energy as
\begin{align}
	\Pi_{\hat\phi\hat\phi}^\eta= \left( \frac{g_e}{e} \right)^2 \frac{m_{\phi}^{2}\,\Pi_{AA}^\eta}{m_{\phi}^{2}-\Pi_{AA}^\eta}\qquad {\rm (vector\,\,DM)}\,.
\end{align}
Scalar and pseudoscalar DM only have one degree of freedom, and therefore Eq.~\eqref{eq:diag_se} takes the form
\begin{align}
	\Pi_{\hat\phi\hat\phi}= \Pi_{\phi\phi}+\sum_\lambda\frac{\Pi_{A\phi}^{\lambda}\Pi_{\phi A}^{\lambda}}{m_\phi^2-\Pi_{AA}^\lambda}\quad {\rm ((pseudo)scalar\;DM)}\,.
\end{align}

As usual, the self-energies appearing in the previous equations are computed from the sum of 1PI diagrams. Working at one loop, there are two graph topologies that can contribute 
\begin{align}
\parbox[c][60pt][c]{120pt}{\centering
	\begin{fmffile}{se1-1loop}
	\begin{fmfgraph*}(70,40)
	\fmfleft{in}
	\fmfright{out}
	\fmf{dashes,tension=2,label=$\overset{Q}{\longrightarrow}$,l.side=left,l.d=3pt}{in,v1}
	\fmf{dashes,tension=2}{v2,out}
	\fmf{fermion,left,tension=0.5}{v1,v2}
	\fmf{fermion,left,tension=0.5}{v2,v1}
	\fmfv{decor.shape=circle,decor.filled=30,decor.size=3thick,label={\footnotesize $\mathcal{O}_1$\;\;},label.angle=-110,l.d=8pt}{v1}
	\fmfv{decor.shape=circle,decor.filled=30,decor.size=3thick,label={\footnotesize \;\;$\mathcal{O}_2$},label.angle=-70,l.d=8pt}{v2}
	\end{fmfgraph*}
	\end{fmffile}
}
\equiv&\; -i\,\bar{\Pi}_{\mathcal{O}_1,\mathcal{O}_2}(Q) \,,\label{eq:graph_topology_1}\\
\parbox[c][80pt][c]{120pt}{\centering
	\begin{fmffile}{se2-1loop}
	\begin{fmfgraph*}(60,80)
	\fmfleft{in}
	\fmfright{out}
	\fmf{dashes,tension=2,label=$\overset{Q}{\longrightarrow}\;\;$,l.side=left,l.d=4pt}{in,v1}
	\fmf{dashes,tension=2}{v1,out}
	\fmf{fermion,right,tension=0.6}{v1,v1}
	\fmfv{decor.shape=circle,decor.filled=30,decor.size=3thick,label={\footnotesize $\mathcal{O}$},label.angle=-90,l.d=8pt}{v1}
	\end{fmfgraph*}
	\end{fmffile}
}
	\equiv&\; -i\,\bar{\Pi}'_{\mathcal{O}}(Q) \,,
	\label{eq:graph_topology_2}
\end{align}
where $\mathcal{O}_{(1,2)}$ is any operator to which the external field, $A$ or $\phi$ (dashed lines), couples. For vector external states these operators carry Lorentz indices that are inherited by $\bar\Pi$ and $\bar\Pi'$.

The full expressions for the self-energies involved in the absorption calculation can be found in Appendix \ref{app:self_ene}. However, as we discuss in the same appendix, due to the absorption kinematics and the hierarchy between the DM and electron velocities, a few diagrams dominate these self-energies. Specifically, we find that the diagonalization of the photon in-medium self-energy (and therefore the derivation of $\Pi_{AA}^\lambda$) reduces to diagonalizing $\bar{\Pi}_{v^i,v^j}$, where the velocity operator is defined by 
\begin{align}\label{eq:diel}
	v^i\equiv\frac{-i\overleftrightarrow{\nabla}_i}{2m_e}\,.
\end{align}
From this it follows that the long wavelength limit of the dielectric function, $\boldsymbol{\varepsilon}(0,\omega)$, which will enter explicitly in the scattering rate calculation, can be derived from $\bar{\Pi}_{v^i,v^j}$:
\begin{align}
\left[ \boldsymbol{\varepsilon} (0,\omega) \right]^{ij} =\mathbb{1}+\frac{\Pi^{ij}_{AA}}{\omega^2}\simeq\mathbb{1}-e^2\frac{\bar{\Pi}_{v^i,v^j}}{\omega^2}\,.
\label{eq:dielectric_notation}
\end{align}
The long wavelength dielectric function, $\boldsymbol{\varepsilon}(0, \omega)$, together with details of its numeric calculation, is reported in Appendix \ref{subapp:numerics_dm_int_rate_convergence}.\footnote{Strictly speaking, the dielectric function is a mixed index tensor, as evident from the defining equation, $J^i = \sigma^i_j A^j = i \omega \left( 1 - \varepsilon^i_j \right) A^j$, where $\varepsilon^i_j = \delta^i_j - \Pi^i_j/\omega^2 = 1+ \Pi^{ij}/\omega^2$ (see the discussion in Appendix A of Ref.~\cite{Coskuner:2019odd} for more details). With a slight abuse of notation we define the matrix $\boldsymbol{\varepsilon}$ which has components $\left[ \boldsymbol{\varepsilon} \right]^{ij} = \varepsilon^i_j$.} For scalar and pseudoscalar DM, the leading order terms in the self-energy of the mostly DM eigenstate are found to be 
\begin{align}
	\Pi_{\hat\phi\hat\phi} \simeq 
	\begin{dcases}
		g_e^2\, \bar{\Pi}_{\bar{v}^2, \bar{v}^2} & \mathrm{(scalar\,\,DM)}\\\\
		g_e^2\frac{\omega^2}{4m_e^2} \bar{\Pi}_{\mathbf{v} \cdot \mathbf{\sigma}, \mathbf{v} \cdot \mathbf{\sigma}}& \mathrm{(pseudoscalar\,\, DM)} 
	\end{dcases}
\end{align}
where we have introduced the operator
\begin{align}
	\bar{v}^2\equiv-\frac{\overleftrightarrow\nabla^2}{8m_e^2}\, .
\end{align}

\begin{figure*}
    \includegraphics[width=0.75\textwidth]{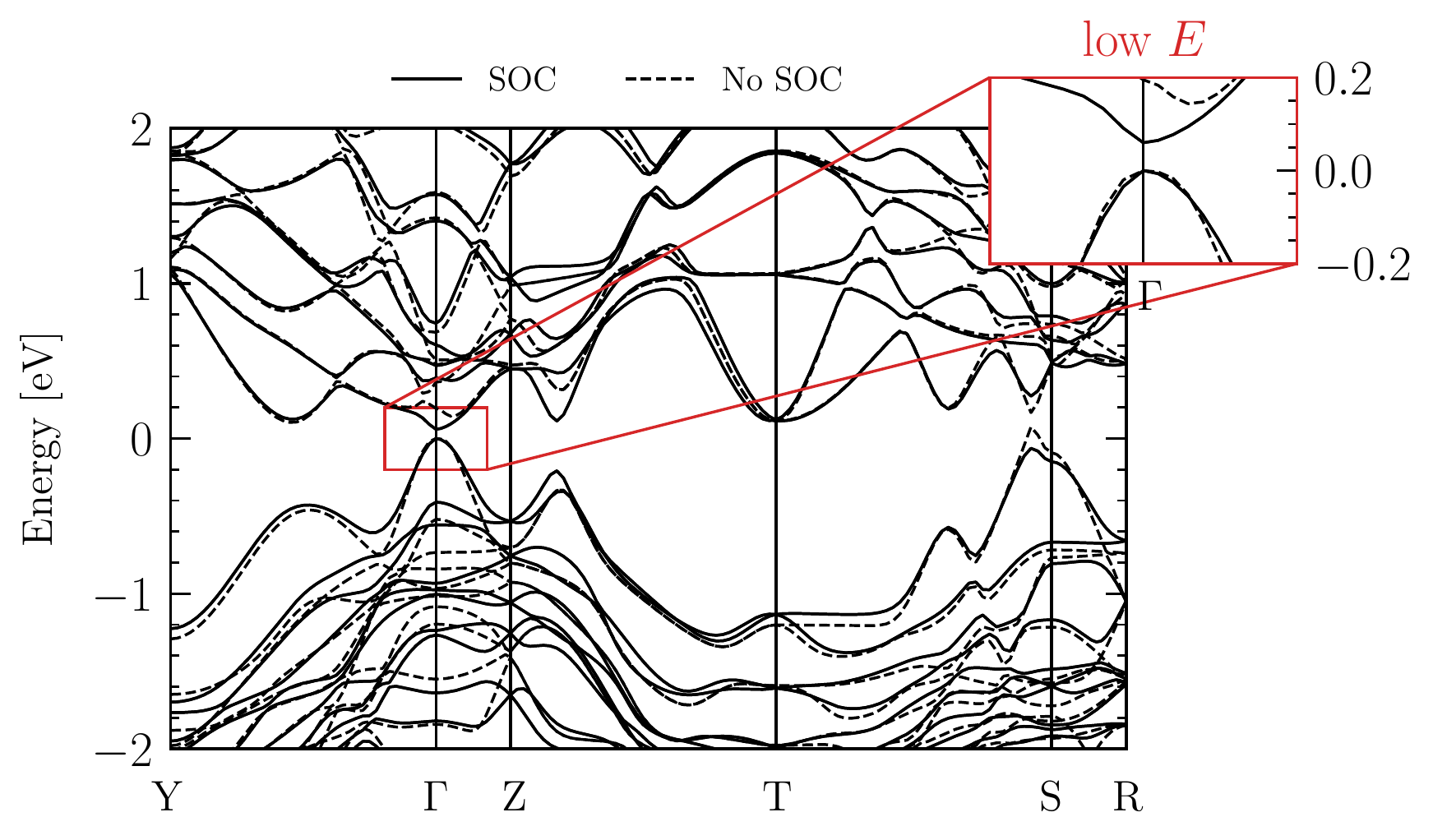}
    \caption{
       \ce{ZrTe5} band structure computed using DFT with SOC (solid lines) and without SOC (dashed lines). The inset highlights the low energy (low $E$) band dispersion, whose details are sampled using a denser $\mathbf{k}$-point grid. The band gap for the SOC band structure is set to the experimental value of $23.5\,\mathrm{meV}$~\cite{Xiong2017}, and the No SOC band structure is shifted accordingly, which gives a larger band gap of $81.6\,\mathrm{meV}$.
        \label{fig:ZrTe5_illustration}
    }
\end{figure*}

\subsection{Scattering}
\label{subsec:formalism_scatter}

In this subsection we proceed to derive the DM scattering rate with spin-dependent electronic wave functions. Generalizing the formulas previously derived in Refs.~\cite{Graham:2012su, Essig:2015cda,Catena:2019gfa, Trickle:2019nya, Liang:2018bdb, Griffin:2021znd}, we can write the DM scattering rate as 
\begin{align}
    \Gamma_{I \rightarrow I'} & = \frac{\pi}{8 V m_e^2 m_\chi^2}  \int d^3q \, \delta \left( E_{I'} - E_I - \omega_{\vec{q}} \right) \nonumber \\
    & \times \overline{\left| \int \frac{d^3k}{(2\pi)^3}\, \widetilde{\Psi}^*_{I'}(\vec{k} + \vec{q}) \cdot \mathbfcal{M}(\vec{q}) \cdot \widetilde{\Psi}_I(\vec{k}) \right|^2}\,
\end{align}
where the bar indicates a spin average (sum) over the incoming (outgoing) DM states, $\widetilde\Psi_I(\vec{k})$ are the Fourier transform of the electronic wave functions defined in Eq.~\eqref{eq:bloch_wf}, and 

\begin{align}
	\omega_{\vec{q}} \equiv \vec{q}\cdot\vec{v}-\frac{q^2}{2m_\chi}\,.
\end{align}

For the SI and SD models of interest here, we can write the free electron scattering amplitude as
\begin{align}
	\mathcal{M}_{ss', \sigma \sigma' }(\vec{q}) = \sqrt{\frac{16 \pi m_\chi^2 m_e^2\overline{\sigma}_e}{\mu_{\chi e}^2}} \frac{f_e}{f_e^0} \,\mathcal{F}_\text{med} \left( \frac{q}{q_0} \right) \mathcal{S}_{ss',\sigma\sigma'}\,,
\end{align}
where $\mathcal{F}_\text{med}\bigl(\frac{q}{q_0}\bigr)$ encodes the momentum dependence induced by the mediator propagator, $f_e(\mathbf{q})/f_e^0$ is the screening factor introduced by in-medium effects, and  $\overline{\sigma}_e$ is a reference cross section defined by
\begin{align}
	\overline{\sigma}_e\equiv\frac{\mu_{\chi e}^2}{64\pi m_\chi^2m_e^2}\sum_{ss',\sigma\sigma'}\left|\mathcal{M}_{ss', \sigma \sigma' }(q_0)\right|^2\,,
\end{align}
with $q_0=\alpha m_e$. The total rate per unit detector mass is then
\begin{widetext}
	\begin{align}
		R = \frac{\pi \overline{\sigma}_e}{V \mu_{\chi e}^2 m_\chi} \frac{\rho_\chi}{\rho_T} \sum_{I, I'} \int \frac{d^3q}{(2\pi)^3} \left( \frac{f_e}{f_e^0} \right)^2 \mathcal{F}_\mathrm{med}^2(q)\, g(\vec{q}, \omega)\,\mathscr{F}_{II'}(\vec{q}) \label{eq:main_scattering_rate} 
	\end{align}
\end{widetext}
where $g(\vec{q}, \omega)$ is the velocity integral defined as 
\begin{align}
	g(\vec{q}, \omega)\equiv\int d^3\vec{v}f_\chi(\vec{v})\,2\pi\delta(\omega-\omega_{\vec{q}})\,,
\end{align}
with $f_\chi(\vec{v})$ being the DM velocity distribution in the laboratory rest frame, which we take to be a boosted Maxwell-Boltzmann distribution with parameters $v_0 = 230\,\mathrm{km}\,\mathrm{s}^{-1}$,
$v_{\rm esc} = 600\,\mathrm{km}\,\mathrm{s}^{-1}$, and $v_e = 240\,\mathrm{km}\,\mathrm{s}^{-1}$.

The crystal form factor $\mathscr{F}_{II'}$ is defined as 
\begin{align}
	\mathscr{F}_{II'}(\vec{q}) & \equiv \overline{\left| \int \frac{d^3k}{(2\pi)^3} \widetilde\Psi^*_{I'}(\vec{k} + \vec{q}) \cdot \mathbfcal{S} \cdot \widetilde\Psi_I(\vec{k})\right|^2}
\end{align}
where the spin operators for the models considered in this work are given by
\begin{align}
	\mathcal{S}_{ss',\sigma\sigma'} = 
	\begin{dcases}
		\delta_{ss'}\delta_{\sigma\sigma'} & \mathrm{(SI)}\,, \\\\
		\frac{1}{\sqrt{3}} \sum_{i} \sigma_{ss'}^i\sigma^i_{\sigma \sigma'}& \mathrm{(SD)}\,,
	\end{dcases}
\end{align}
and $\sigma_{i}$ are the Pauli matrices. 
Given these expressions the form factors, $\mathscr{F}_{II'}$, for the SI and SD models take the form,\footnote{The absence of the overall factor of two, relative to the SI rate formula given in Ref.~\cite{Griffin:2021znd}, can be understood from the sum over the states. If the wave functions are spin independent then $\sum_{IF} \rightarrow \sum_{IF} \sum_{ss'}$, where $s$ ($s'$) indexes the initial (final) electron spin state. These spin sums contribute the extra factor of two, bringing Eq.~\eqref{eq:main_scattering_rate} and the rate formula in Ref.~\cite{Griffin:2021znd} into agreement.}
\begin{align}
	\mathscr{F}_{II'} = 
	\begin{dcases}
		|\mathcal{T}_{II'}|^2 & \mathrm{(SI)}\,, \\
		\frac{1}{3} \,\mathbfcal{T}_{II'}^* \cdot \mathbfcal{T}_{II'} & \mathrm{(SD)} \,,
	\end{dcases}
\end{align}
where we have defined the DM model independent transition form factors, $\mathcal{T}_{II'}$ and $\mathbfcal{T}_{II'}$, as
\begin{align}
	\mathcal{T}_{II'} & = \int \frac{d^3k}{(2\pi)^3} \widetilde{\Psi}_{I'}^*(\vec{k} + \vec{q}) \cdot \widetilde{\Psi}_I(\vec{k})\,, \label{eq:T_I_Ip_SI}\\
	\mathbfcal{T}_{II'} & = \int \frac{d^3k}{(2\pi)^3} \widetilde{\Psi}_{I'}^*(\vec{k} + \vec{q}) \cdot \boldsymbol{\sigma} \cdot \widetilde{\Psi}_I(\vec{k})\,.
\end{align}

\section{Detection Rates in \texorpdfstring{$\ce{ZrTe5}$}{TEXT}}
\label{sec:ZrTe5_results}

We will now apply the formalism developed in the previous section to our benchmark SOC target: \ce{ZrTe5}. The band structure of \ce{ZrTe5}, with and without the inclusion of SOC effects, is shown in Fig.~\ref{fig:ZrTe5_illustration}. The details of the DFT calculation can be found in Appendix~\ref{subapp:numerics_dft}. The dominant effect of SOC is to shift the valence and conduction bands closer at the $\Gamma$ point relative to the No SOC calculation. 

While in theory the calculation of DM interaction rates is identical for $\mathcal{O}(\si{\meV})$ and $\mathcal{O}(\si{\eV})$ gap semiconductors, in practice one must be careful about sampling the 1BZ. This is because these $\mathcal{O}(\si{\meV})$ energy differences generally only occur in small volumes within the 1BZ. To account for this we sample the 1BZ with a higher $\mathbf{k}$-point density in regions corresponding to the low energy band structure. For \ce{ZrTe5} this occurs near the $\Gamma$ point, and we split the phase space in to two separate regions, ``low $E$" and ``high $E$", which we describe now. The low $E$ region consists of the highest two (one) valence bands and lowest two (one) conduction bands, for the calculation with (without) SOC, sampled on a ``mini-BZ" grid. This mini-BZ grid is a rescaled uniform Monkhorst-Pack grid~\cite{Monkhorst:1976zz}; each $\mathbf{k}$ is scaled by a factor of $1/5$, giving a 125$\times$ $\mathbf{k}$-point sampling in that region. This region will give the dominant contribution to absorption of DM with mass $\lesssim 100\,\si{\meV}$, as well as DM scattering via a light mediator. The high $E$ region includes all the bands outside the low $E$ region, sampled with a standard Monkhorst-Pack uniform grid. The DM absorption rates in Sec.~\ref{subsec:ZrTe5_results_abs} are a combination of the low and high $E$ regions. The DM scattering rates in Sec.~\ref{subsec:ZrTe5_results_scatter} will be shown for both regions, and it will be clear when one dominates the other.

We compute the Bloch wave functions, Eq.~\eqref{eq:bloch_wf}, in both regions within the framework of DFT with {\sc Quantum ESPRESSO}~\cite{Giannozzi2009, Giannozzi2017, Giannozzi2020}; details can be found in Appendix~\ref{subapp:numerics_dft}. The DM absorption and scattering rates are computed with an extended version of \href{https://exceed-dm.caltech.edu}{\textsf{EXCEED-DM}}~\cite{EDC2021, Griffin:2021znd} which includes the formalism developed in Sec.~\ref{sec:dm_interaction_rate_formalism}, and is publicly available on Github \href{https://github.com/tanner-trickle/EXCEED-DM}{\faGithub}. 

For each of the models considered in the following subsections, we will show the projected constraints from three different calculations. The curves labelled ``SOC" are computed with the inclusion of SOC effects, the curves labelled ``No SOC" do not include any SOC effects, and those labelled ``Partial SOC" are a combination of the SOC and No SOC calculations, obtained using the energy levels computed with SOC, and the wave functions without SOC. While the Partial SOC results are not a consistent calculation, they aid in understanding how much of the difference between the SOC and No SOC results is due to the changes in the band structure versus the inclusion of the spin dependent wave functions. Generally we find that the changes in the band structure are more influential than the spin dependence in the wave functions, but the latter can still be important.  

Lastly, we note that previous works \cite{Coskuner:2019odd, Hochberg:2017wce, Geilhufe:2019ndy} have derived excitation rates analytically by exploiting the putative Dirac nature of \ce{ZrTe5}. While a direct comparison to assess the validity of the analytic approximations is dubious since we do not observe a conical band structure (see Appendix~\ref{subapp:numerics_dft}), previous estimates from Ref.~\cite{Coskuner:2019odd} are shown in Figs.~(\ref{fig:abs_reach},~\ref{fig:scatter_reach}). In App.~\ref{app:an_rates} we discuss the validity of these analytic approximations in more general Dirac materials.

\subsection{Absorption}
\label{subsec:ZrTe5_results_abs}

\begin{figure*}[ht]
	\centering
    \includegraphics[width=\textwidth]{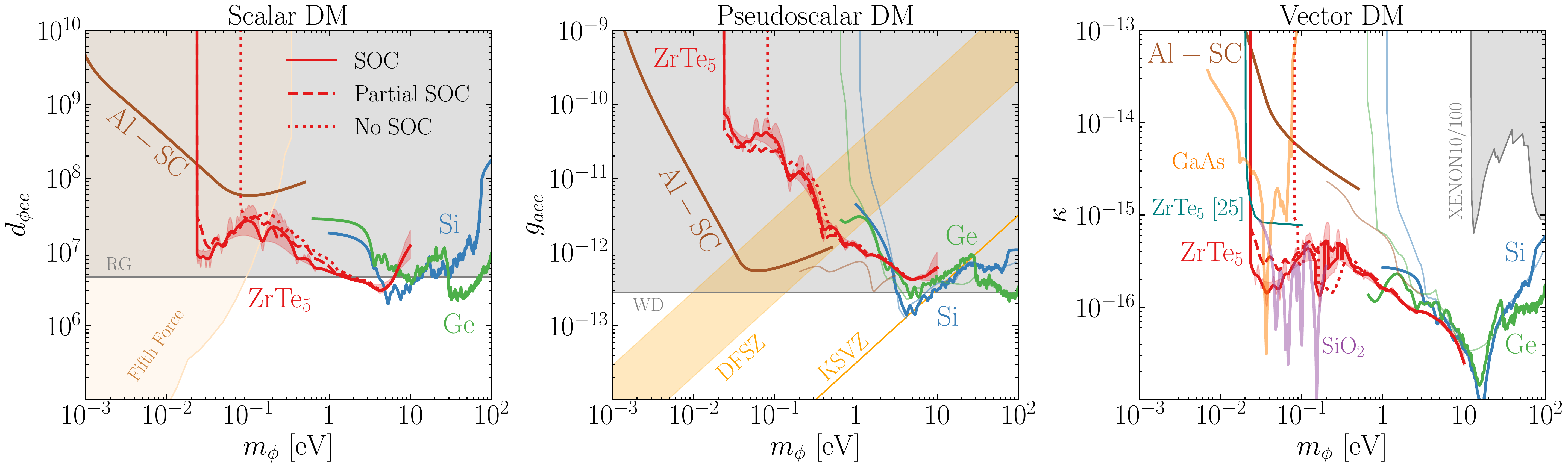}
	\caption{\label{fig:abs_reach}
    Comparison of projected $95\%$ C.L. reach (3 events, no background) assuming one kg-year exposure for scalar (left), pseudoscalar (center) and vector (right) DM. We compare our results with (solid) and without (dotted) SOC for electronic absorption in a \ce{ZrTe5} target (red), with the ones for semiconductor silicon (Si, blue) and germanium (Ge, green) targets \cite{Mitridate:2021ctr}, superconducting aluminum (Al-SC, brown) \cite{Mitridate:2021ctr}), phononic absorption in polar materials \cite{Griffin:2018bjn, Knapen:2021bwg} (\ce{GaAs} in orange and \ce{SiO2} in purple), and previous estimates for \ce{ZrTe5} (teal) \cite{Coskuner:2019odd}. We also show the projected constraints combining the SOC energy levels with the No SOC wave functions, (``Partial SOC", red, dashed) to explicitly show the effect of the spin dependent wave functions. Constraints are expressed in terms of the commonly adopted parameters shown in Eq.~\eqref{eq:g_e_reparam}. Shaded red bands correspond to different parameterizations of the electron width $\delta \in [10^{-1.5}, 10^{-0.5}] \, \omega$ used in calculating the self-energies (see  e.g.\ Eq.~\eqref{eq:diagram1}), with the solid line corresponding to $\delta = 10^{-1} \omega$. Thin lines indicate results obtained by rescaling the optical data. Also shown are the direct detection limits from XENON10/100~\cite{Bloch:2016sjj}, fifth force constraints \cite{Adelberger:2003zx}, and stellar cooling constraints from red giants (RG) \cite{Hardy:2016kme}, and white dwarfs (WD) \cite{MillerBertolami:2014rka}. For the pseudoscalar scenario we also report the couplings corresponding to the QCD axion in KSVZ and DFSZ models, for $0.28\leq \tan\beta\leq140$~\cite{ParticleDataGroup:2018ovx}.}
\end{figure*}

For the models considered, Eq.~\eqref{eq:abs_lagr}, our results are shown in Fig.~\ref{fig:abs_reach}. For ease of comparison we map the constraints on the $g_e$ parameters in Eq.~\eqref{eq:abs_lagr} to a more commonly used notation,
\begin{align}
    g_e = \begin{cases}
        \frac{4\pi m_e}{M_\text{Pl}}d_{\phi ee} & \mathrm{(scalar)} \\
        g_{aee} & \mathrm{(pseudoscalar)} \\
        e \kappa & \mathrm{(vector)} \, ,
    \end{cases}
    \label{eq:g_e_reparam}
\end{align}
where $M_\text{Pl} = 1.22\times10^{19}\,\si{\GeV}$ is the Planck mass. 

For all the benchmark models, the inclusion of SOC effects dominantly impacts the low mass reach where the SOC corrections to the band structure are most relevant. Most notably, the lowest testable DM mass is shifted as a consequence of the different band gaps: $23.5\,\si{\meV}$ with SOC, and $81.6\,\si{\meV}$ without SOC. At higher masses the SOC effects are milder and, as expected, the SOC reach approaches the reach without SOC effects. The close agreement between the ``Partial SOC" and ``SOC" curves indicates that changes to the energy levels are what is mainly driving the difference in the ``SOC" and ``No SOC" calculations.  

For the scalar and vector DM models we find that \ce{ZrTe5} is superior at low DM masses relative to a superconducting aluminum target, another target material with an $\mathcal{O}(\si{\meV})$ gap ($0.6\,\mathrm{meV}$ for the Al-SC curves shown here). However for pseudoscalar DM, for $m_\phi \lesssim \si{\eV}$, Al-SC yields better sensitivity than \ce{ZrTe5}. This can be attributed to the large amount of screening present in the vector DM case (but not in the pseudoscalar DM case) for Al-SC. 

Shaded bands correspond to different width parameterizations, $\delta \in [10^{-1.5}, 10^{-0.5}] \, \omega$. In theory, the absorption rate calculation is independent of the choice of width; however, when sampling the 1BZ discretely this is not the case, and practically the goal is to find results that have a weak dependence on this parameter. The discrepancy in the shaded bands should be viewed as an uncertainty in the calculation. The constraints turn up on the left hand side because of the band gap, and on the right hand side because of the finite number of bands used in the calculation. All bands for which $E - E_F < 4\,\mathrm{eV}$, where $E_F$ is the valence band maximum were included; see Appendix~\ref{subapp:numerics_dm_int_rate_convergence} for more details.

\subsection{Scattering}
\label{subsec:ZrTe5_results_scatter}

\begin{figure*}[t]
	\centering
    \includegraphics[width=\textwidth]{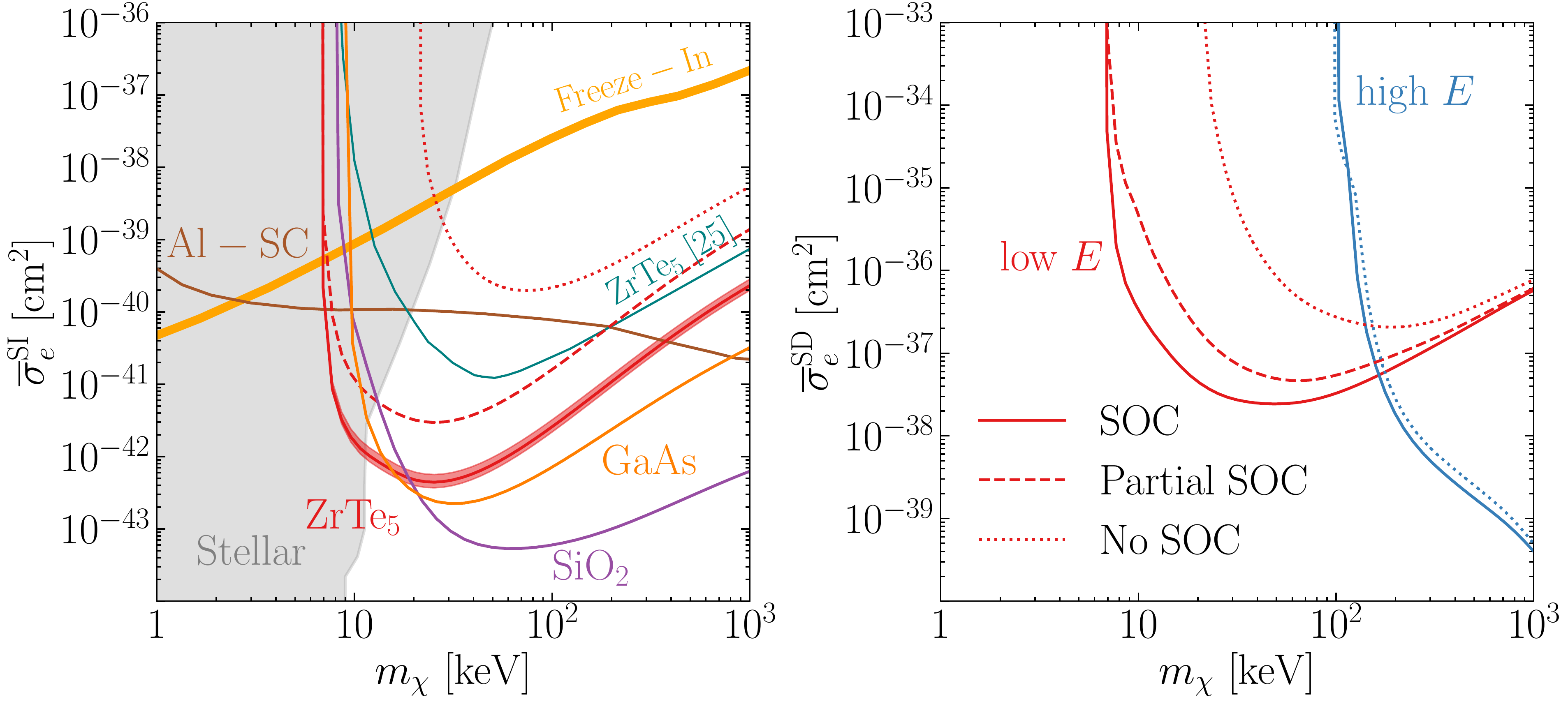}
	\caption{\label{fig:scatter_reach}
        Projected constraints on DM-electron scattering cross sections at the 95\% C.L. (three events, no background) assuming one kg-year exposure for two benchmark models shown in Eq.~\eqref{eq:scatt_lagr}. \textbf{Left}: SI model with a light mediator ($\mathcal{F}_\text{med} = (q_0/q)^2$), screened with the static dielectric shown in Fig~\ref{fig:dielectric}. The red solid (dashed) curve shows the constraints with (without) the inclusion of SOC effects. For comparison we also show projected constraints from single phonon excitations in \ce{GaAs} (orange) and \ce{SiO2} (purple) computed with \textsf{PhonoDark}~\cite{Trickle:2020oki} (assuming an energy threshold of $\omega_\text{min} = 20\,\mathrm{meV}$), electronic excitations in an aluminum superconductor~\cite{Hochberg2021} (brown), and previous estimates for \ce{ZrTe5} (teal)~\cite{Coskuner:2019odd}. We also show the projected constraints combining the SOC energy levels with the No SOC wave functions, (``Partial SOC", red, dashed) to explicitly show the effect of the spin dependent wave functions. Stellar constraints (gray) are taken from Ref.~\cite{Vogel:2013raa} and the freeze-in benchmark (orange) is taken from Ref.~\cite{Dvorkin:2019zdi}. \textbf{Right}: SD model with a heavy mediator ($\mathcal{F}_\text{med} = 1$). Curves labelled ``low/high $E$" include transitions restricted to the low/high $E$ regions discussed in Sec.~\ref{sec:ZrTe5_results}.}
\end{figure*}

We now consider DM-electron scattering in \ce{ZrTe5} for the two benchmark models, standard SI and standard SD interactions, shown in Eq.~\eqref{eq:scatt_lagr}. Specifically we consider a light mediator for the SI model and a heavy mediator for the SD model. For the SI model a light mediator was chosen due to its high sensitivity to the lowest energy excitations, as well as for ease of comparison with other proposals which commonly report constraints on this model. The SD model was chosen to highlight the effect of spin dependent wave functions.\footnote{In the SD model, $\mathcal{F}_\text{med} = 1$ also for a light mediator due to the dominance of longitudinal component. Here we focus on the heavy mediator case. To avoid perturbativity constraints on the couplings, $g_\chi g_e \lesssim (4\pi)^2$, one needs $m_{A'} \lesssim 3\,\mathrm{GeV} \left( \frac{ 10^{-37}\si{\square\centi\metre}}{\overline{\sigma}^\text{SD}_e} \right)^{1/4}$ for $\si{\keV} < m_\chi < \si{\MeV}$.}

The results are shown in Fig.~\ref{fig:scatter_reach} and we discuss them in detail here. Constraints computed in this work are shown in red, with shaded bands corresponding to the uncertainty in the calculation of the screening factor/dielectric function from the electron width parameter, discussed previously in Sec.~\ref{subsec:ZrTe5_results_abs}. 

When considering the SI model with a light mediator we include anisotropic screening effects in the $f_e(\mathbf{q})/f_e^0 = (\hat{\mathbf{q}} \cdot \boldsymbol{\varepsilon}(\mathbf{q}, \omega) \cdot \hat{\mathbf{q}})^{-1}$ factor. $\boldsymbol{\varepsilon}(\mathbf{q}, \omega)$ is the dielectric tensor, and this screening factor is especially important for the sub-MeV DM masses considered here. Since in this model the scattering rate is dominated by events with small $q$, we approximate $\boldsymbol{\varepsilon}(\mathbf{q}, \omega) \approx \boldsymbol{\varepsilon}(0, \omega)$, such that we replace the dielectric with the anisotropic, long wavelength dielectric function shown in Fig.~\ref{fig:dielectric}. 

For the SI model, we find that the contribution from transitions in the low $E$ region, discussed earlier in Sec.~\ref{sec:ZrTe5_results}, dominate the scattering rate. Therefore, in the left panel of Fig.~\ref{fig:scatter_reach}, we only show the results derived from transitions within the low $E$ region. For the massive mediator SD model, in the right panel of Fig.~\ref{fig:scatter_reach}, we see that the low $E$ contributions dominate at small DM masses. However for $m_\chi \gtrsim 100\,\mathrm{keV}$, when the high $E$ contributions at $\mathcal{O}(100\,\mathrm{meV})$ become kinematically available, the high $E$ contributions are dominant. This is due to the fact that when scattering via a heavy mediator the rate is no longer dominated by the smallest momentum transfers. While we did not explicitly include transitions between the low and high $E$ regions, we note that these are only expected to be important for masses where the reach is comparable between the regions, and will not affect the conclusions.

We find that, for the SI model with a light mediator, the inclusion of SOC effects significantly alters the reach for the whole DM mass range considered since the rate is dominated by small energy/momentum depositions. For the SD model with a massive mediator the SOC effects are most prominent for low DM masses when the scattering is probing the band structure near the band gap, which is the most affected by SOC effects. We also see that at the lowest masses the ``Partial SOC" curve is closer to the ``SOC" than the ``No SOC" lines. This shows that while the change to the energy levels is the dominant effect when including SOC, the spin dependence of the wave functions can give $\mathcal{O}(1)$ variations.

The left hand side of all the constraint curves are determined by the band gap. The smallest kinematically allowed DM mass is $m_\chi = 6\,\mathrm{keV}$ for the SOC calculation with $E_g = 23.5\,\mathrm{meV}$, and $m_\chi = 21\,\mathrm{keV}$ for the No SOC with $E_g = 81.6\,\mathrm{meV}$. As mentioned in Sec.~\ref{subsec:ZrTe5_results_abs}, we only consider bands up to $4\,\mathrm{eV}$ above the valence band maximum. Kinematically this means that we are only including all contributions for $m_\chi < \si{\MeV}$, and explains why our projections stop there.

\section{Conclusions}
\label{sec:conclusions}

Materials with strong spin-orbit coupling, such as \ce{ZrTe5}, are promising targets in which electronic excitations can be utilized to search for sub-MeV DM. Their $\mathcal{O}(\si{\meV})$ band gaps lead to sensitivity to new DM parameter space via both absorption and scattering processes, without relying on detecting single collective excitation modes.

However, due to the spin-orbit coupling, in these materials the electron spin is no longer a good quantum number, and the spin sums over electronic states cannot be trivially reduced. This introduces interesting wrinkles in the DM absorption and scattering rate calculations, which we extended to account for these effects. In addition, we updated the \href{https://exceed-dm.caltech.edu}{\textsf{EXCEED-DM}} program~\cite{EDC2021, Griffin:2021znd}, which computes DM-electron interaction rates from first principles, to be compatible with this input for future study of general targets with spin-orbit coupling.

We considered a wide range of DM models and processes to which materials with SOC are sensitive: absorption of vector, pseudoscalar, and scalar DM in Sec.~\ref{subsec:ZrTe5_results_abs}, and scattering via heavy and light mediators via spin-independent and spin-dependent scattering potentials in Sec.~\ref{subsec:ZrTe5_results_scatter}. We found that for sub-eV vector and scalar DM absorption, \ce{ZrTe5} is a far superior target relative to an aluminum superconductor. We also found more optimistic projections for SI scattering via a light mediator than previous estimates, and computed, for the first time, the projected constraints on an SD model with a heavy mediator. Our projections for \ce{ZrTe5} lay the foundation for further first-principles studies of materials with strong spin-orbit coupling as targets in direct detection experiments.

\begin{acknowledgments}
    H-Y.C. and M.B. were supported by the National Science Foundation under Grant No.~DMR-1750613. A.M., T.T., K.Z.\ and Z.Z.\ were supported by the U.S. Department of Energy, Office of Science, Office of High Energy Physics, under Award No.~DE-SC0021431, and the Quantum Information Science Enabled Discovery (QuantISED) for High Energy Physics (KA2401032). K.Z. was also supported by a Simons Investigator Award. Z.Z.\ was also supported by the U.S. Department of Energy under the grant DE-SC0011702. The computations presented here were conducted in the Resnick High Performance Computing Center, a facility supported by Resnick Sustainability Institute at the California Institute of Technology.
\end{acknowledgments}

\appendix

\onecolumngrid

\section{Numerical Details}
\label{app:numerics}

\subsection{Density Functional Theory (DFT)}
\label{subapp:numerics_dft}

The DFT calculations are carried out within the generalized gradient approximation (GGA) \cite{perdew1996generalized} using the {\sc Quantum Espresso} code \cite{giannozzi2009quantum} with and without spin-orbit coupling (SOC) included. We use the \ce{ZrTe5} experimental lattice constants, $a=3.9797$ \AA, $b=14.470$ \AA, and $c=13.676$ \AA ~of the orthorhombic crystal structure \cite{fjellvaag1986structural}. We employ fully relativistic pseudopotentials for calculations including SOC, and scalar relativistic pseudopotentials for calculations without SOC, in both cases generated with Pseudo Dojo \cite{perdew1981self,troullier1991efficient,van2018pseudodojo}. In each case, we use a 3265 eV kinetic energy cutoff on a uniform $4\times 4 \times 2$ Brillouin zone (BZ) grid to compute the electron density. To systematically converge the absorption and scattering rates, for the high $E$ region we compute the electronic wave functions with 200, 300, and 400 eV cutoffs on $10\times10\times10$, $12\times12\times12$, and $14\times 14 \times 14$ $\mathbf{k}$-grids. For the low $E$ region, we compute the wave functions with 650, 750, and 850 eV cutoffs on $8\times8\times8$, $9\times9\times9$, and $10\times 10 \times 10$ uniform $\mathbf{k}$-grids in a small reciprocal-space volume that includes the low-energy band dispersion. The convergence of these calculations is discussed in Appendix~\ref{subapp:numerics_dm_int_rate_convergence}.
\\
\indent
The computed band structure of \ce{ZrTe5} is presented in Fig.~\ref{fig:ZrTe5_illustration}, where we correct the band gap with a scissor shift to match the experimental band gap for the calculation with SOC. The inset shows in detail the dispersion near the band edges, highlighting the linear dispersion along the intralayer directions $\Gamma$-Y and $\Gamma$-Z. Note that in interlayer directions (not shown in Fig.~\ref{fig:ZrTe5_illustration}) the dispersion is not linear or conical. This band structure obtained by combining the experiment lattice constant and the Perdew-Burke-Ernzerhof (PBE) exchange correlation functional is consistent with a previous study \cite{fan2017transition}. While the presence of a Dirac cone in $\rm ZrTe_5$ is still under debate~\cite{Zheng2016, Chen2015, Liu2018, Li:2014bha, Chen2015a, Yuan2016, Chen2017, Monserrat2019, Wu2016, Nair2017, Zhang2017, Moreschini2016}, pursuing more extensive tests of crystal structure and DFT functionals, or carrying out beyond-DFT band structure calculations, is beyond the scope of this work.

\subsection{DM Interaction Constraint Convergence and Dielectric Function}
\label{subapp:numerics_dm_int_rate_convergence}

\begin{figure}[t]
	\centering
    \includegraphics[width=\textwidth]{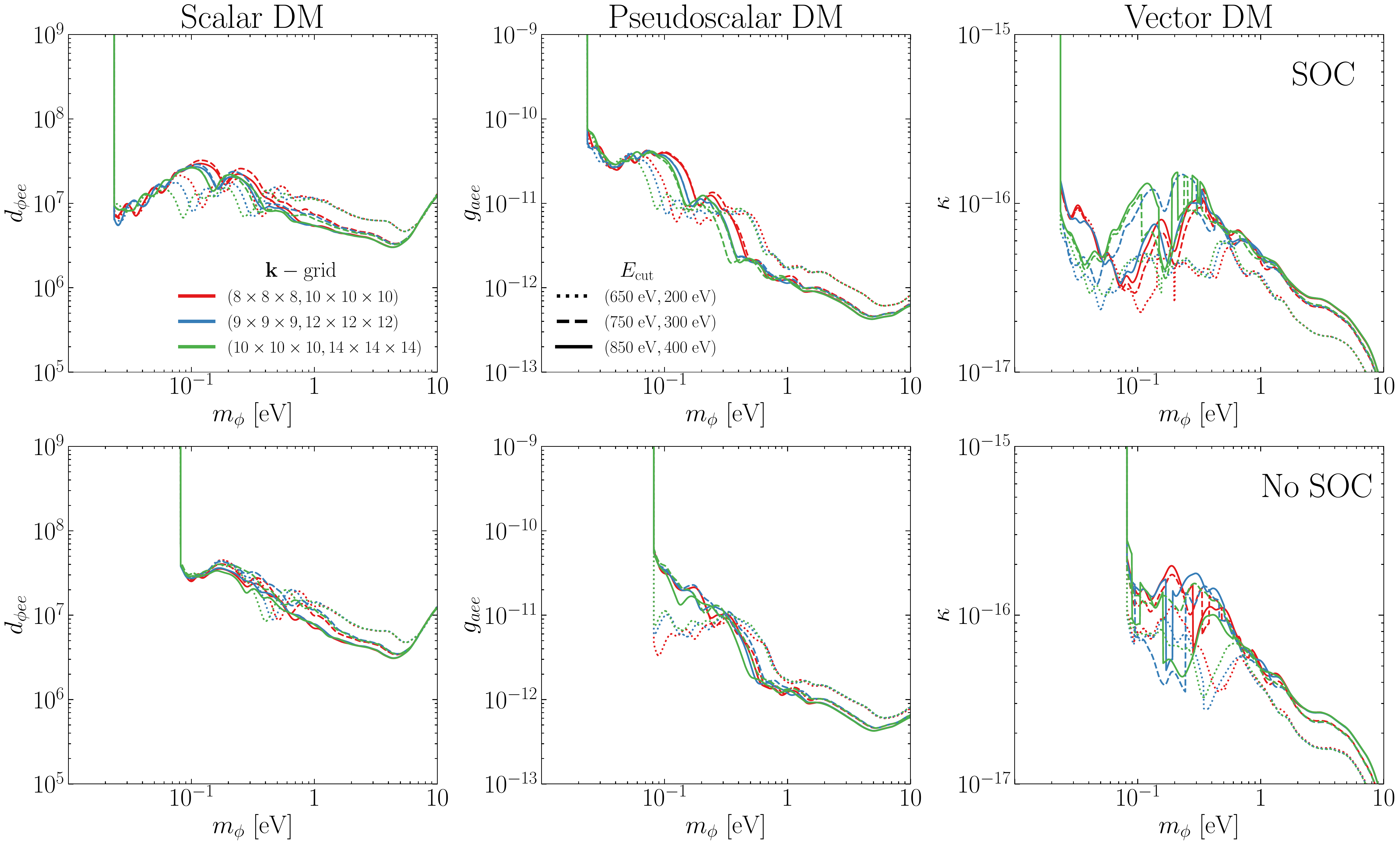}
	\caption{Convergence of the constraints on DM absorption, for the models discussed in Sec.~\ref{subsec:ZrTe5_results_abs}, with respect to the $\mathbf{k}$ point sampling ($\mathbf{k}-$grid) and plane wave energy cutoff, $E_\text{cut}$. The first row includes SOC effects while the second row does not. Absorption rates were computed by adding the contributions from the low $E$ and high $E$ regions, and the first (second) value in the legends corresponds to the parameter used in the low (high) $E$ calculation. For example, the red dotted line corresponds to a calculation in which the low (high) $E$ region was sampled on an $8\times8\times8\; (10 \times 10 \times 10)$ Monkhorst-Pack grid in the 1BZ, with $E_\text{cut} = 650\; (200) \,\si{\eV}$. All curves assume a width parameter of $\delta = 10^{-1} \omega$.}
    \label{fig:abs_reach_convergence}
\end{figure}

\begin{figure}[t]
	\centering
    \includegraphics[width=\textwidth]{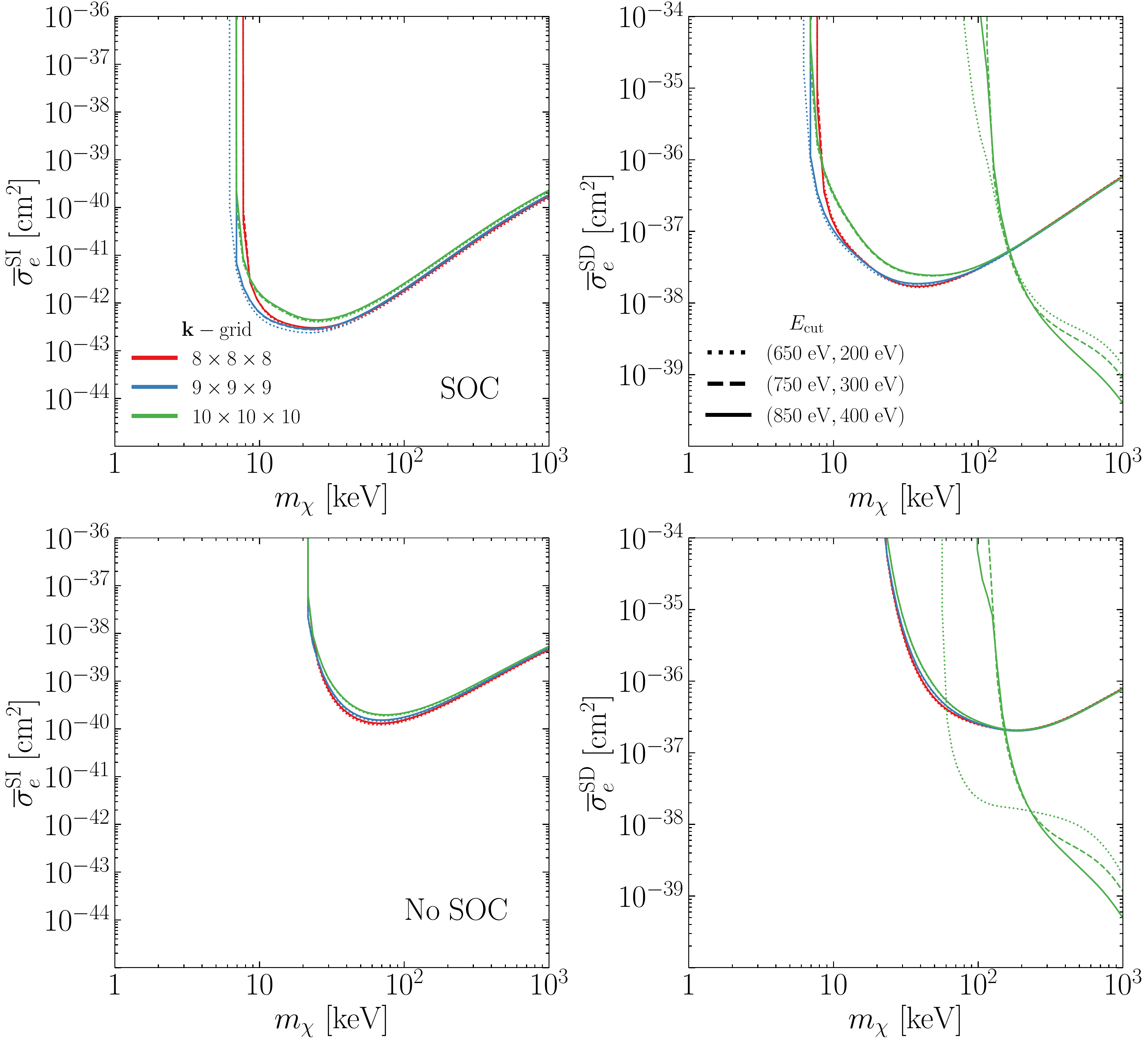}
	\caption{Convergence of the constraints on DM scattering, for the models discussed in Sec.~\ref{subsec:ZrTe5_results_scatter}, with respect to the $\mathbf{k}$ point sampling ($\mathbf{k}$-grid) and plane wave energy cutoff, $E_\text{cut}$. The first row includes SOC effects while the second row does not. The collection of constraints dominant at the lowest masses corresponds to the low $E$ transitions, and the other set corresponds to the high $E$ transitions. Similar to Fig.~\ref{fig:abs_reach_convergence}, the $E_\text{cut}$ parameters in the legend correspond to the values used for the low/high $E$ regions.}
    \label{fig:scatter_reach_convergence}
\end{figure}

In this appendix we will discuss some details of the DM scattering and absorption rate calculations, as well as the long wavelength, anisotropic dielectric function, $\boldsymbol{\varepsilon}(0, \omega)$. Since the main focus of this paper is the effect of SOC, only the electronic wave functions near the Fermi surface are needed. This is because, in \ce{ZrTe5}, SOC effects are approximately $\mathcal{O}(10\,\mathrm{meV})$, and therefore a very small perturbation for states $> \si{\eV}$ away from the Fermi surface. We are therefore safely within the ``valence to conduction" regime, discussed in more detail in Ref.~\cite{Griffin:2021znd}, and do not need to study deeper, core electronic levels, or larger energy states where the electrons are close to free. DFT is the preferred tool for studying these transitions, and the two main convergence parameters are the number of $\mathbf{k}$-points in the 1BZ sampling, and the plane wave expansion cutoff, $E_\text{cut}$. In both the low $E$ and high $E$ regions we sample $\mathbf{k}$ points uniformly with a Monkhorst-Pack grid. The only difference is that the low $E$ points are scaled by $1/5$ relative to the high $E$ region. Convergence of the DM absorption and scattering constraints with respect to the $\mathbf{k}$ point sampling and $E_\text{cut}$ parameters are shown in Fig.~\ref{fig:abs_reach_convergence} and Fig.~\ref{fig:scatter_reach_convergence} respectively. The constraints in the main text are identical to the most converged constraints shown in Figs.~(\ref{fig:abs_reach_convergence},~\ref{fig:scatter_reach_convergence}). Generally we see faster convergence with respect to $E_\text{cut}$ than the $\mathbf{k}$ point density, and slightly faster convergence for the DFT calculation which omits SOC effects than those which include them. We also note that all-electron reconstruction effects were omitted here since we are focusing on very small DM masses, and therefore kinematically limited to small $q$ transitions. However these effects could be important for studies of DM scattering in \ce{ZrTe5} at higher masses, or for larger experimental thresholds.

\begin{figure}[t]
    \includegraphics[width=0.7\textwidth]{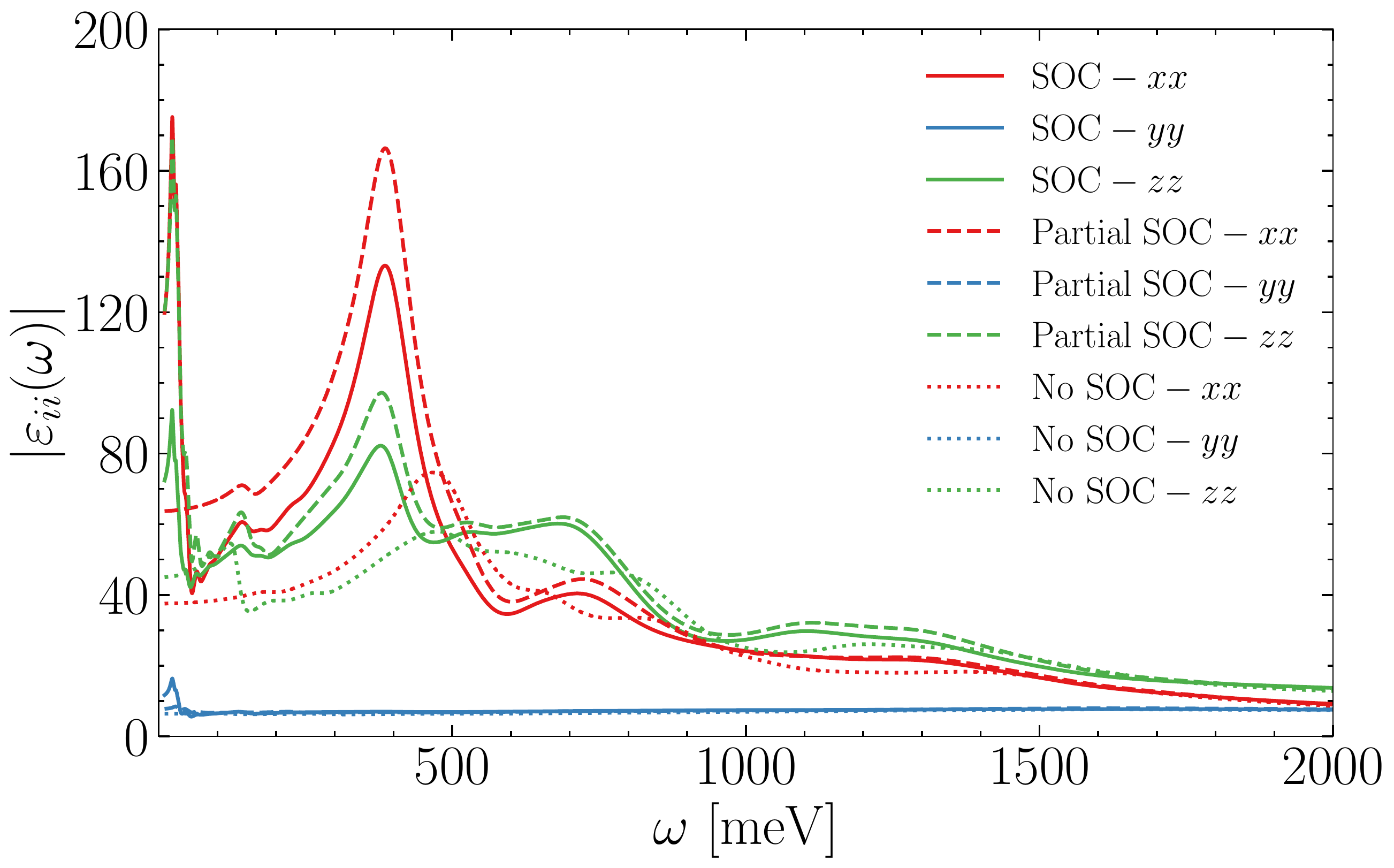}
    \caption{Magnitude of the dielectric function of \ce{ZrTe5} computed with SOC (solid), without SOC (dotted), and a combination of the calculations with and without SOC (dashed), as described in Sec.~\ref{sec:ZrTe5_results}. The directional dependence of the dielectric function is due to the anisotropic nature of \ce{ZrTe5}. Note that while non-local corrections are not included in this figure we found they have a small $\mathcal{O}(10\%)$ effect. These results are obtained with an electronic broadening of $\delta = 10^{-1} \omega$.}
    \label{fig:dielectric}
\end{figure}

The dielectric function in the long wavelength limit is shown in Fig.~\ref{fig:dielectric}, was used as an intermediate to compute a few different constraints. Specifically it was used to screen the SI scattering rate, and it can be shown that the vector DM absorption rate, as well as the pseudoscalar DM absorption rate when wave functions are spin independent, can be related to the dielectric function. Moreover this calculation serves as a useful benchmark to compare future DFT calculations. 

\section{Generalized Self-Energies}
\label{app:self_ene}
In this appendix we provide the expressions for the self-energies used in the main text, namely $\Pi_{AA}$, $\Pi_{A\phi}$, and $\Pi_{\phi\phi}$. Since electrons in the target are non-relativistic, we will work in the framework of NR EFT closely following Ref. \cite{Mitridate:2021ctr}, generalizing the results to anisotropic materials with sizable SOC. 

At leading order in the NR EFT, it can be shown \cite{Mitridate:2021ctr} that the electron-photon coupling reads:
\begin{align}
	\label{eq:L_psiA_eff}
	\mathcal{L}_{\psi A}^\text{eff} = -e\, A_0\, \psi_+^\dagger \psi_+ 
	-\frac{ie}{2m_e} \,\vec{A} \cdot \left(\psi_+^\dagger \overleftrightarrow{\nabla} \psi_+\right)
	+\frac{e}{2m_e}\,(\nabla\times\vec{ A}) \cdot \left(\psi_+^\dagger \,\vec{\Sigma}\, \psi_+ \right)
	-\frac{e^2}{2m_e}\,\vec{A}^2\,\psi_+^\dagger \psi_+
\,,
\end{align}
where $\vec{\Sigma}={\rm diag}(\vec{\sigma},\vec{\sigma})$, and $\psi_+=\frac{1}{2}(1+\gamma^0)\psi_{\rm NR}$ with $\psi_{\rm NR}$ being the NR electron field defined as
\begin{align}
	\psi(\vec{x},t)=e^{-im_et}\,\psi_{\rm NR}(\vec{x},t)\,.
\end{align}
For vector DM, by simply replacing $e A^\mu\to eA^\mu-g_e\phi^\mu$ in eq.~\eqref{eq:L_psiA_eff}, we obtain:
\begin{align}
\mathcal{L}_\text{int}^\text{eff} =\; g\,\phi_0\, \hat\psi_+^\dagger \hat\psi_+ 
+\frac{ig}{2m_e} \,\vec{\phi} \cdot \left(\hat\psi_+^\dagger \overleftrightarrow{\nabla} \hat\psi_+\right)
-\frac{g}{2m_e}\,(\nabla\times\vec{\phi}) \cdot \left(\hat\psi_+^\dagger \,\vec{\Sigma}\, \hat\psi_+ \right)
+\frac{ge}{m_e}\,\vec{\phi}\cdot\vec{A}\,\hat\psi_+^\dagger \hat\psi_+
-\frac{g^2}{2m_e}\,\vec{\phi}^2\,\hat\psi_+^\dagger \hat\psi_+&
\\
\text{(vector DM).}&\nonumber
\end{align}
In deriving the effective interaction Lagrangian for scalar and pseudoscalar DM, we have to keep some NLO terms in the NR expansion. This is because, as discussed in \cite{Mitridate:2021ctr}, the LO order terms contain factors of the momentum transfer, $q$, which in the absorption limit induces a larger suppression compared to the electron velocity. Therefore, keeping all the NLO order terms that do not contain factors of $q$ we obtain
\begin{align}
	\mathcal{L}_\text{int}^\text{eff} = 
	\begin{cases}
		g\,\phi\, \hat\psi_+^\dagger \hat\psi_+ 
		+\frac{g}{8m_e^2}\,\phi \left(\hat\psi_+^\dagger \overleftrightarrow{\nabla}^2 \hat\psi_+ \right)
		-\frac{ige}{2m_e^2}\,\phi\,\vec{A}\cdot \left(\hat\psi_+^\dagger \overleftrightarrow{\nabla} \hat\psi_+\right)
		& \text{(scalar DM)}\,,\\[8pt]
		-\frac{g}{2m_e}\, (\nabla\phi)\cdot \hat\psi_+^\dagger \,\vec{\Sigma}\,\hat\psi_+  
		+\frac{ig}{4m_e^2}\,(\partial_t\phi) \left(\hat\psi_+^\dagger\,\vec{\Sigma}\cdot \overleftrightarrow{\nabla} \,\hat\psi_+\right)
		& \text{(pseudoscalar DM)}\,.
	\end{cases}
\label{eq:L_int_eff_scalar}
\end{align}

With these effective interactions, we are now ready to derive the expressions for the self-energies. By using the photon-electron coupling given in eq.~\eqref{eq:L_psiA_eff}, we obtain the expression for $\Pi_{AA}^{\mu\nu}$ in terms of the loop diagrams $\Pi_{\mathcal{O}_1\mathcal{O}_2}$ and $\Pi'_\mathcal{O}$ defined in eq.~\eqref{eq:graph_topology_1} and \eqref{eq:graph_topology_2}:
\begin{align}
	\Pi_{AA}^{00}&=-e^2\bar{\Pi}_{\mathbb{1}\mathbb{1}}\label{eq:pi_00}\\
	\Pi_{AA}^{i0}&=-e^2\left(\bar{\Pi}_{v^i\mathbb{1}}+\hlr{\frac{iq^l}{2m_e}\epsilon_{iml}\,\bar{\Pi}_{\sigma^m\mathbb{1}}}\right)\\
	\begin{split}
	\Pi_{AA}^{ij}&=-e^2\left[\bar{\Pi}_{v^iv^j}+\hlr{\frac{iq^l}{2m_e}\left(\epsilon_{jml}\,\bar{\Pi}_{v^i\sigma^m}+\epsilon_{iml}\,\bar{\Pi}_{\sigma^mv^j}\right)} - \frac{q^lq^r}{4m_e^2}\epsilon_{ilm}\epsilon_{jrn}\bar{\Pi}_{\sigma^m\sigma^n}\right]-\omega_p^2\delta^{ij}\label{eq:pi_ij}
	\end{split}
\end{align}
where $\omega_p=\sqrt{\frac{n_e e^2}{m_e}}$ is the plasma frequency, and we have highlighted in \hlr{red} terms that vanish in absence of sizable spin-orbit coupling (in this specific case, they vanish because ${\rm tr}[\vec{\sigma}^i]=0$ in absence of SOC). Since vector DM couples to electrons in the same way of the photon but with a rescaled coupling, $\kappa = g_e/e$, we have 
\begin{align}
	\Pi_{\phi\phi}^{\mu\nu}=-\kappa\Pi_{\phi A}^{\mu\nu}=\kappa^2\Pi_{AA}^{\mu\nu}\hspace{10em}\text{(vector DM)}\,.
\end{align}

For scalar DM, by using the interactions given in eq.~\eqref{eq:L_int_eff_scalar}, we get 
\begin{align}
	\Pi^0_{\phi A}&=-g_e e\Big(\bar{\Pi}_{\mathbb{1}\mathbb{1}}-\bar{\Pi}_{\mathbb{1}\bar v^2}+\hlr{ \frac{iq^k}{4m_e}\epsilon_{ijk}\bar{\Pi}_{\mathbb{1}\tilde v^{ij}}}\Big)\hspace{20em}\text{(scalar DM)}  \label{eq:scalar_se1}\\
	\Pi^i_{\phi A}&=-g_e e\left[\bar{\Pi}_{\mathbb{1}v^i}-\bar{\Pi}_{\bar{v}^2v^i} +\hlr{\frac{iq^k}{4m_e}\epsilon_{ljk}\bar{\Pi}_{\tilde{v}^{lj}v^i}}+\hlr{\frac{iq^l}{4m_e}\epsilon_{lim}\left(\bar{\Pi}_{\mathbb{1}\sigma^m}-\bar{\Pi}_{\bar{v}^2\sigma^m}\right)}+\hlb{\frac{iq^k}{4m_e}\frac{iq^r}{4m_e}\epsilon_{ijk}\epsilon_{mlr}\bar{\Pi}_{\tilde{v}^{ij}\sigma^m}}+\frac{1}{m_e}\bar{\Pi}'_{v^i}\right]\label{eq:scalar_se2}\\
	\Pi_{\phi\phi}&=g_e^2\left[\bar{\Pi}_{\mathbb{1}\mathbb{1}}-\bar{\Pi}_{\mathbb{1}\bar v^2}-\bar{\Pi}_{\bar v^2\mathbb{1}} + \bar{\Pi}_{\bar v^2\bar v^2} + \hlr{\frac{iq^k}{4m_e}\epsilon_{ijk}\left(\bar{\Pi}_{\mathbb{1}\tilde v^{ij}}+\bar{\Pi}_{\tilde v^{ij}\mathbb{1}}+\bar{\Pi}_{\bar{v}^2\tilde v^{ij}}+\bar{\Pi}_{\tilde v^{ij}\bar{v}^2}\right)}+\hlb{\frac{iq^k}{4m_e}\frac{iq^r}{4m_e}\epsilon_{ijk}\epsilon_{mlr}\bar{\Pi}_{\tilde{v}^{ij}\tilde{v}^{ml}}}\right]\,,
	\label{eq:scalar_se3}
\end{align}
where we have introduced the operator $\tilde v^{ij}\equiv \sigma^i v^j$, and as before highlighted in \hlr{red} the terms that vanish in absence of sizable SOC. The terms highlighted in \hlb{blue}, instead, vanish in isotropic materials without SOC.

Similarly, by using the couplings given in eq.~\eqref{eq:L_int_eff_scalar}, we derive the expression for the self-energies of pseudoscalar DM:
\begin{align}
	\Pi_{\phi A}^0&=\hlr{-ig_e e\left(\frac{q^i}{2m_e}\bar{\Pi}_{\sigma^i\mathbb{1}}-\frac{\omega}{2m_e}\bar{\Pi}_{\tilde v^{ii}\mathbb{1}}\right)}\\
	\Pi_{\phi A}^i&=g_e e\left(\hlr{\frac{q^j}{2m_e}\bar{\Pi}_{\sigma^j v^i}-\frac{\omega}{2m_e}\bar{\Pi}_{\tilde v^{jj}v^i}-\frac{q^j}{2m_e}\frac{q^l}{2m_e}\epsilon_{lir}\bar{\Pi}_{\sigma^j\sigma^r}+\bar{\Pi}'_{\sigma^i}}\right)\hspace{5em}\text{(pseudoscalar DM)}\\
	\Pi_{\phi\phi}&=-g_e^2\left[\frac{q^i}{2m_e}\frac{q^j}{2m_e}\bar{\Pi}_{\sigma^i\sigma^j}-\frac{q^i}{2m_e}\frac{\omega}{2m_e}\left(\bar{\Pi}_{\sigma^i\tilde v^{jj}}+\bar{\Pi}_{\tilde v^{jj}\sigma^i}\right)+\frac{\omega^2}{4m_e^2}\bar{\Pi}_{\tilde v^{ii}\tilde v^{jj}}\right]\,.
\end{align} 

\begin{table}[t!]
  \renewcommand{\arraystretch}{1}
  \setlength\extrarowheight{-10pt}
  \begin{tabular}{c@{\hskip 0.1in}|@{\hskip 0.1in}c@{\hskip 0.3in}c@{\hskip 0.3in}c@{\hskip 0.3in}c}
	\toprule
		$\bar{\Pi}_{\mathcal{O}_1\mathcal{O}_2}$		&	$\mathbb{1}$		&	$\sigma$ 						&	$v,\;\tilde v$				&	$\bar v^2$ 					\Tstrut\Bstrut\\ \midrule
		$\mathbb{1}$						&	$v_e^2v_\phi^2$	&	$\dfrac{m_\phi}{m_e} v_\phi^2$		&	$v_e^2v_\phi$				&	$\dfrac{m_\phi}{m_e}v_e^2v_\phi^{2}$					\Tstrut\Bstrut\\ 
		$\sigma$							&					&	$1$							&	$\dfrac{m_\phi}{m_e}v_\phi$	&	$v_e^2$						\Tstrut\Bstrut\\ 
		$v,\;\tilde v$						& 					&								&	$ v_e^2$					&	$\dfrac{m_\phi}{m_e}v_e^2v_\phi$ 	\Tstrut\Bstrut\\ 
		$\bar{v}^2$						&					&								&							& 	$v_e^4$						\Tstrut\Bstrut\\ \bottomrule
\end{tabular}
\caption{Self-energies scaling with the DM and electron velocities in the absorption limit. Notice that each insertion of the identity operator induces a suppression of order $v_e v_\phi$ due to the wave-function orthogonality, and that parity odd self-energies receive an additional suppression of order $q/k$.}\label{tab:self_scaling}
\end{table} 
Due to the absorption kinematics ($q\sim m_\phi v_\phi\ll\omega\sim m_\phi$), and the hierarchy that exists between the DM velocity, $v_\phi\sim10^{-3}$, and the electrons' typical velocity in a crystal, $v_e\sim 10^{-2}$; only a few terms are actually relevant in the self-energy expressions given above. To facilitate the following discussion, in Table \ref{tab:self_scaling} we summarize the velocity scaling of all the terms appearing in the self-energy expression given above. By using these scaling relations it is easy to see that the photon self-energy (and therefore also the DM self-energy) is dominated by its spatial components, specifically by the $\bar{\Pi}_{v^iv^j}$ term. For scalar DM, $\Pi_{\phi\phi}$ is dominated by the term $\bar{\Pi}_{\bar v^2\bar v^2}$, and the mixing self-energies $\Pi_{A\phi}$ are suppressed by one power of $v_\phi$. Finally, for pseudoscalar DM, $\Pi_{\phi\phi}$ is dominated by $\bar{\Pi}_{\tilde v\tilde v}$ and the mixing self-energies are again suppressed.

So far we have ignored the tadpole terms $\bar{\Pi}'_{\mathcal{O}}$. They can be written in terms of the electronic wave functions as \cite{Mitridate:2021ctr}:
\begin{align}
\bar{\Pi}'_{\mathcal{O}}=-\frac{1}{V}\sum_{I}f_{I}\langle I|\mathcal{O}|I\rangle\,,
\end{align}
and are usually related to macroscopic quantities of the material. Specifically, $\bar{\Pi}'_{v_i}$ and $\bar{\Pi}'_{\sigma_i}$ are related to the current and spin densities of the material (which both vanish for the case of \ce{ZrTe5}). For the case of \ce{ZrTe5}, the only non-vanishing tadpole term is $\bar{\Pi}'_{\mathbb{1}}=n_e$, which enters in the expression for the vector self-energy. However, we never explicitly compute this term. Instead, we exploit the relation 
\begin{align}
\bar{\Pi}_{\mathbb{1}\mathbb{1}}^{ij}=\frac{m_e^2}{\omega^2}\left(\bar{\Pi}_{v^iv^j}-\frac{\delta^{ij}}{m_e}\bar{\Pi}'_{\mathbb{1}}\right)
\end{align}
where $\bar{\Pi}_{\mathbb{1}\mathbb{1}}=\frac{q^i}{m_e}\bar{\Pi}_{\mathbb{1}\mathbb{1}}^{ij}\frac{q^j}{m_e}$. Indeed, as discussed in \cite{PhysRevB.95.155203}, a direct numerical derivation of $\bar{\Pi}_{v^iv^j}-\frac{\delta^{ij}}{m_e}\bar{\Pi}'_{\mathbb{1}}$ would be affected by numerical errors in the $\omega\to 0$ limit. 

Let's conclude this section by discussing more in detail the scaling relations given in Table \ref{tab:self_scaling}. Indeed, while some of them are trivial, others require some explanation. The expression for the loop diagrams $\bar{\Pi}_{\mathcal{O}_1,\mathcal{O}_2}$ is given by \cite{Mitridate:2021ctr}:
\begin{align}
	-i\,\bar{\Pi}_{\mathcal{O}_1,\mathcal{O}_2} =
	\frac{i}{V}\, \sum_{I'I}\, \frac{f_{I'}-f_I}{\omega-\omega_{I'I}+i\delta_{I'I}^{}} 
	\, \langle I' | \,\mathcal{O}_1 \,e^{i\vec{q}\cdot\vec{x}} | I\rangle \langle I | \,\mathcal{O}_2 \,e^{-i\vec{q}\cdot\vec{x}} | I'\rangle \,,
\label{eq:diagram1}
\end{align}
where $V$ is the total volume,  $\omega_{I'I}\equiv E_I'-E_I$, $\delta_{I'I}\equiv\delta\,{\rm sgn}(\omega_{I'I})$, and  $f_I$, $f_I'$ are the occupation numbers (which, at zero temperature, equal one for states below the Fermi surface, and zero for states above it). From this expression we can see that self-energies involving the identity operators contain the matrix element  $\langle i',\vec{k}'| \,e^{i\vec{q}\cdot\vec{x}} \,| i, \vec{k}\rangle$ which vanishes in the $q\to 0$ limit since $|i',\vec{k}'\rangle$ and $|i,\vec{k}\rangle$ are distinct energy eigenstates and therefore orthogonal.
At $\mathcal{O}(q)$, we have $\langle i',\vec{k}'| \,e^{i\vec{q}\cdot\vec{x}} \,| i, \vec{k}\rangle\simeq i\vec{q}\cdot \langle i',\vec{k}'| \,\vec{x} \,| i, \vec{k}\rangle$. 
One way to compute this matrix element is to trade the position operator for the momentum operator via its commutator with the Hamiltonian. Here we will assume that the Hamiltonian has the form $H=\frac{\vec{p}^2}{2m_e} + V(\vec{x})$, ignoring the possibility of momentum-dependent or non-local terms in the potential. While these terms can introduce mild corrections ($\mathcal{O}(10\%)$) we do not expect them to change the overall scaling of the self-energies so we can ignore them in this context. With this assumption in mind, we can write the matrix element involving the position operator as 
\begin{equation}
\langle i',\vec{k}'| \,\vec{x} \,| i, \vec{k}\rangle 
= -\frac{1}{E_{i',\vec{k}'} -E_{i,\vec{k}}} \langle i',\vec{k}'| \,[\vec{x}, H] \,| i, \vec{k}\rangle 
= -\frac{i}{m_e(E_{i',\vec{k}'} -E_{i,\vec{k}})} \langle i',\vec{k}'| \,\vec{p} \,| i, \vec{k}\rangle \,.
\label{eq:trick}
\end{equation}
Writing the wave functions in the Bloch form, we find:
\begin{equation}
\langle i',\vec{k}'| \,e^{i\vec{q}\cdot\vec{x}} \,| i, \vec{k}\rangle 
=\delta_{\vec{k}',\vec{k}} \,\frac{\vec{q}}{m_e\, \omega_{i'i,\,\vec{k}}} \cdot \sum_{\vec{G}}  (\vec{k}+\vec{G}) \,\left(u^s_{i', \vec{k}, \vec{G}}\right)^* \,u^s_{i, \vec{k},\vec{G}} +\mathcal{O}(q^2)\,.
\end{equation}
where $\omega_{i'i,\,\vec{k}}\equiv E_{i',\vec{k}} - E_{i,\vec{k}}$. Therefore, in the absorption limit, each identity operator entering in a self-energy diagram induces a suppression of order $v_ev_\phi$.

Parity-odd self energies also vanish in the $q\to0$ limit. Let's show this explicitly for the case of  $\bar{\Pi}_{v^i\bar{v}^2}$. By rewriting the electronic wave function in the Bloch form, we can write $\bar{\Pi}_{v^i\bar{v}^2}$ as
\begin{align}
	\Pi_{\bar v^2 v^i}=\frac{1}{V}\frac{1}{16m_e^3}\sum_{\substack{i'\in\,\text{con.}\\ i\,\in\,\text{val.}}}\sum_{\vec{k}}
	\left(\frac{\big[\sum_{\vec{G}}(2\vec{k}+\vec{G}+\vec{q})u^{s*}_{i'\vec{k}\vec{G}}u^s_{i\vec{k}\vec{G}}\big]\big[\sum_{\vec{G}}(2\vec{k}+\vec{G}+\vec{q})^2u^{\lambda*}_{i'\vec{k}\vec{G}}u^\lambda_{i\vec{k}\vec{G}}\big]}{\omega-\omega_{i'i,\vec{k},\vec{k}+\vec{q}}+i\delta_{i'i,\vec{k},\vec{k}+\vec{q}}}-i\to i'\right)\,.
\end{align}
By parity invariance the Bloch coefficients satisfy the relation $u_{i\vec{k}\vec{G}}^s=u_{i-\vec{k}-\vec{G}}^s$, therefore, at order $q^0$ we have $\bar{\Pi}_{v^i\bar{v}^2}= -\bar{\Pi}_{v^i\bar{v}^2}=0$. The first non-vanishing contribution arises at order $q$ and is given by 
\begin{align}
	\bar{\Pi}_{\bar v^2 v^i}=\frac{1}{V}\frac{q^i}{4m_e^3}\sum_{\substack{i'\in\,\text{con.}\\ i\,\in\,\text{val.}}}\sum_{\vec{k}}
	\left(\frac{\big[\sum_{\vec{G}}u^{s*}_{i'\vec{k}+\vec{q}\vec{G}}u^s_{i\vec{k}\vec{G}}\big]\big[\sum_{\vec{G}}(2\vec{k}+\vec{G})^2u^{\lambda*}_{i'\vec{k}\vec{G}}u^\lambda_{i\vec{k}\vec{G}}\big]}{\omega-\omega_{i'i,\vec{k}}+i\delta_{i'i,\vec{k}}}-i\to i'\right)\,.
\end{align}
Therefore, instead of the naive $v_e^3$ scaling, $\bar{\Pi}_{\bar v^2v^i}$ scales as $(m_\phi/m_e) v_e^2v_\phi$ in the absorption limit. By following an analogous derivation, we can conclude that any parity-odd operator receives an additional $\frac{m_\phi v_\phi}{m_e v_e}\sim \frac{q}{k}$ suppression respect to its naive scaling. 

\section{Analytic Approximations in Dirac Materials}
\label{app:an_rates}

Dirac materials are defined by having a conical band structure near the Fermi surface. They are ``Dirac" since the electronic dispersion relation in this conical region is linear in $k$, similar to the solutions to the Dirac equation,
\begin{align}
	\left(i \slashed{\partial} - m\right)\psi(x)=0
\end{align}
describing free fermions. The presence of this conical structure in Dirac materials implies the existence of low energy excitations which satisfy a rescaled version of this equation,
\begin{align}
	\left(i \widetilde{\slashed{\partial}} - \Delta \right)\psi(x)=0\,,
    \label{eq:rescaled_Dirac_eq}
\end{align}
where $2\Delta$ is a band gap between the two cones, and $\widetilde{\partial}_\mu \equiv (\partial_t, v_F^x \partial_x, v_F^y \partial_y,v_F^z \partial_z)$, with $\vec{v}_F$ the directionally dependent Fermi velocity. The solutions to Eq.~\eqref{eq:rescaled_Dirac_eq} can be found analytically, and most previous works~\cite{Geilhufe:2019ndy, Hochberg:2017wce, Coskuner:2019odd} studying DM-electron interactions in 3D Dirac materials used these analytic solutions as the Bloch wave functions in Eq.~\eqref{eq:bloch_wf}. Specifically, they used these analytic wave functions to derive scattering and absorption rates.

However, the subtlety is that solutions to Eq.~\eqref{eq:rescaled_Dirac_eq} cannot be the electronic Bloch wave functions since they are not eigenstates of the crystal Hamiltonian, $H = \vec{p}^2/2m_e + V$. Therefore while the excitations which satisfy the rescaled Dirac equation, Eq.~\eqref{eq:rescaled_Dirac_eq}, are certainly related to the electronic Bloch wavefunctions, they are, generally, not the appropriate wave functions to use when computing DM interaction rates.

To further illustrate this point we will briefly discuss the most well known Dirac material, graphene. Even though it is only two dimensional it will serve as a good example to illustrate the difference between the electronic Bloch wave functions and those which satisfy the Dirac equation. Our discussion here will closely follow Ref.~\cite{Lozovik2008}, to which we refer the reader for further details. 

Graphene has two carbon atoms within a unit cell which form a hexagonal lattice structure. The Bloch wave functions, satisfying the crystal Hamiltonian, are typically found using the ``tight-binding" method, which assumes that the Bloch wave functions are a linear combination of the atomic wave wave functions of each of the carbon atoms,
\begin{align}
    \Psi_{i, \mathbf{k}}(\mathbf{x}) = \sum_{j = A, B} \psi_{j, \mathbf{k}}(\mathbf{x}) X_{j, \mathbf{k}}(\mathbf{x}) \label{eq:total_Bloch}
\end{align}
where the``A" and ``B" indexes refer to the individual carbon atoms (equivalently the individual carbon atom sublattices), $\psi_{j, \mathbf{k}}$ are some coefficient functions and $X_{j, \mathbf{k}}$ are the specific linear combination of the atomic wave functions which forms a Bloch state,
\begin{align}
    X_{j, \mathbf{k}}(\mathbf{x}) = \frac{1}{\sqrt{N}} \sum_{\mathbf{r}} e^{i \mathbf{k} \cdot \mathbf{r}} \psi^\text{atom}_{j}(\mathbf{x} - \mathbf{r} - \mathbf{r}^0_j) \, .
    \label{eq:core_electron}
\end{align}
Here $\mathbf{r}$ is a lattice vector, $\mathbf{r}^0_j$ is the equilibrium position of the carbon atom on the $j^\text{th}$ sublattice, and $N$ is the number of unit cells in the lattice. The Bloch nature of the $X_{j, \mathbf{k}}$ functions can be seen explicitly by noticing that $X_{j, \mathbf{k}}(\mathbf{x} + \mathbf{r}) = e^{i \mathbf{k} \cdot \mathbf{r}} X_{j, \mathbf{k}}(\mathbf{x})$. Assuming that the $\psi_{j, \mathbf{k}}$'s are lattice periodic implies that $\Psi_{i, \mathbf{k}}$ is also a valid Bloch state.
The idea behind this decomposition is that the $\psi_{j, \mathbf{k}}$'s are slowly varying functions, or envelope functions, in position space, while the atomic wave functions contain the high frequency behavior, being very localized to the atomic sites. Using this intuition we can simplify the full Sch\"odinger equation near the Dirac point
\begin{align}
   \left( -\frac{\nabla^2}{2m_e} + V(\mathbf{x}) - E_{i, \mathbf{k}} \right) \Psi_{i, \mathbf{k}} = 0
\end{align}
to
\begin{align}
    0 = & \sum_{j = A, B} -\frac{1}{m_e} \nabla \psi_{j, \mathbf{k}} \cdot \nabla X_{j, \mathbf{k}} + \psi_{j, \mathbf{k}} \left( -\frac{\nabla^2}{2m_e} + V - E_{i, \mathbf{k}} \right) X_{j, \mathbf{k}} \, . \label{eq:Schrodinger_expanded}
\end{align}
This equation can now be ``coarse-grained" by integrating out the pieces close to the center of the atoms with the operator, $\int_{\Omega_l} d^3\mathbf{x} \,  X_{l, \mathbf{k}}^*$ for both $l \in \{ A, B \}$ sublattices. Assuming that $\psi$ varies slowly over these regions, we can pull $\psi_{j, \mathbf{k}}$ out of these integrals and Eq.~\eqref{eq:Schrodinger_expanded} becomes two equations,
\begin{align}
    0 = \sum_{j = A, B} \left( - \frac{1}{m_e} \langle X_{l, \mathbf{k}} | \nabla | X_{j, \mathbf{k}} \rangle \cdot \nabla - \delta_{l, j} E_{i, \mathbf{k}} \right) \psi_{j, \mathbf{k}} \label{eq:tight-binding-approx}
\end{align}
for each $l = A, B$, where the expectation value of $-\nabla^2/2m_e + V$ with respect to $X_{i, \mathbf{k}}$ vanishes since we are implicitly assuming $\mathbf{k}$ is close to the Dirac point, i.e. at the peak of the conical band.. From symmetry arguments it can be shown that $\langle X_{A, \mathbf{k}} | \nabla | X_{B, \mathbf{k}} \rangle \propto \hat{\mathbf{x}} - i \hat{\mathbf{y}}$ and therefore Eq.~\eqref{eq:tight-binding-approx} can be further simplified to,
\begin{align}
    v_F \left( \sigma \cdot \mathbf{k} \right) \begin{pmatrix} \psi_{A, \mathbf{k}} \\ \psi_{B, \mathbf{k}} \end{pmatrix} = E_{i, \mathbf{k}} \begin{pmatrix} \psi_{A, \mathbf{k}} \\ \psi_{B, \mathbf{k}} \end{pmatrix}
\end{align}
which is exactly the rescaled Dirac equation, with $v_F$ the Fermi velocity parameter.

Therefore we see that the $\psi_{i, \mathbf{k}}$ components of the total Bloch wave functions in Eq.~\eqref{eq:total_Bloch} are what satisfies the Dirac equation, not the $\Psi_{i, \mathbf{k}}$ which should be used in the excitation rate calculations. Moreover note that the $\sigma$ operator does not act in spin-space but rather in ``sublattice" space, and therefore for spin-dependent excitation rates the spin dependence follows from the $X_{j, \mathbf{k}}$ functions. 

There are circumstances where the analytic expressions can be used as as approximation. If the tight-binding approximation is valid, and the Bloch wave functions can be cleanly separated in to high and low momentum components (as was just done for graphene), then for $q$ much smaller than typical momentum scale of the $X$ functions the spin independent transition form factors, e.g., Eq.~\eqref{eq:T_I_Ip_SI} if $\Psi$ is spin-independent, can reduce to the previously used analytic expressions. In these targets the agreement between an analytic and numeric approach is then indicative of how good the tight-binding approximation is. However not all Dirac cones necessarily appear from the same tight-binding approximation as in graphene, and a detailed study of the Bloch wave functions, along with the band structure, should be done to understand whether any analytic approximations will be valid.

\bibliographystyle{apsrev4-1}
\bibliography{bibliography}

\begin{thebibliography}{68}%
\makeatletter
\providecommand \@ifxundefined [1]{%
 \@ifx{#1\undefined}
}%
\providecommand \@ifnum [1]{%
 \ifnum #1\expandafter \@firstoftwo
 \else \expandafter \@secondoftwo
 \fi
}%
\providecommand \@ifx [1]{%
 \ifx #1\expandafter \@firstoftwo
 \else \expandafter \@secondoftwo
 \fi
}%
\providecommand \natexlab [1]{#1}%
\providecommand \enquote  [1]{``#1''}%
\providecommand \bibnamefont  [1]{#1}%
\providecommand \bibfnamefont [1]{#1}%
\providecommand \citenamefont [1]{#1}%
\providecommand \href@noop [0]{\@secondoftwo}%
\providecommand \href [0]{\begingroup \@sanitize@url \@href}%
\providecommand \@href[1]{\@@startlink{#1}\@@href}%
\providecommand \@@href[1]{\endgroup#1\@@endlink}%
\providecommand \@sanitize@url [0]{\catcode `\\12\catcode `\$12\catcode
  `\&12\catcode `\#12\catcode `\^12\catcode `\_12\catcode `\%12\relax}%
\providecommand \@@startlink[1]{}%
\providecommand \@@endlink[0]{}%
\providecommand \url  [0]{\begingroup\@sanitize@url \@url }%
\providecommand \@url [1]{\endgroup\@href {#1}{\urlprefix }}%
\providecommand \urlprefix  [0]{URL }%
\providecommand \Eprint [0]{\href }%
\providecommand \doibase [0]{http://dx.doi.org/}%
\providecommand \selectlanguage [0]{\@gobble}%
\providecommand \bibinfo  [0]{\@secondoftwo}%
\providecommand \bibfield  [0]{\@secondoftwo}%
\providecommand \translation [1]{[#1]}%
\providecommand \BibitemOpen [0]{}%
\providecommand \bibitemStop [0]{}%
\providecommand \bibitemNoStop [0]{.\EOS\space}%
\providecommand \EOS [0]{\spacefactor3000\relax}%
\providecommand \BibitemShut  [1]{\csname bibitem#1\endcsname}%
\let\auto@bib@innerbib\@empty
\bibitem [{\citenamefont {Essig}\ \emph
  {et~al.}(2012{\natexlab{a}})\citenamefont {Essig}, \citenamefont {Mardon},\
  and\ \citenamefont {Volansky}}]{Essig:2011nj}%
  \BibitemOpen
  \bibfield  {author} {\bibinfo {author} {\bibfnamefont {R.}~\bibnamefont
  {Essig}}, \bibinfo {author} {\bibfnamefont {J.}~\bibnamefont {Mardon}}, \
  and\ \bibinfo {author} {\bibfnamefont {T.}~\bibnamefont {Volansky}},\ }\href
  {\doibase 10.1103/PhysRevD.85.076007} {\bibfield  {journal} {\bibinfo
  {journal} {Phys. Rev.}\ }\textbf {\bibinfo {volume} {D85}},\ \bibinfo {pages}
  {076007} (\bibinfo {year} {2012}{\natexlab{a}})},\ \Eprint
  {http://arxiv.org/abs/1108.5383} {arXiv:1108.5383 [hep-ph]} \BibitemShut
  {NoStop}%
\bibitem [{\citenamefont {Graham}\ \emph {et~al.}(2012)\citenamefont {Graham},
  \citenamefont {Kaplan}, \citenamefont {Rajendran},\ and\ \citenamefont
  {Walters}}]{Graham:2012su}%
  \BibitemOpen
  \bibfield  {author} {\bibinfo {author} {\bibfnamefont {P.~W.}\ \bibnamefont
  {Graham}}, \bibinfo {author} {\bibfnamefont {D.~E.}\ \bibnamefont {Kaplan}},
  \bibinfo {author} {\bibfnamefont {S.}~\bibnamefont {Rajendran}}, \ and\
  \bibinfo {author} {\bibfnamefont {M.~T.}\ \bibnamefont {Walters}},\ }\href
  {\doibase 10.1016/j.dark.2012.09.001} {\bibfield  {journal} {\bibinfo
  {journal} {Phys. Dark Univ.}\ }\textbf {\bibinfo {volume} {1}},\ \bibinfo
  {pages} {32} (\bibinfo {year} {2012})},\ \Eprint
  {http://arxiv.org/abs/1203.2531} {arXiv:1203.2531 [hep-ph]} \BibitemShut
  {NoStop}%
\bibitem [{\citenamefont {Lee}\ \emph {et~al.}(2015)\citenamefont {Lee},
  \citenamefont {Lisanti}, \citenamefont {Mishra-Sharma},\ and\ \citenamefont
  {Safdi}}]{Lee:2015qva}%
  \BibitemOpen
  \bibfield  {author} {\bibinfo {author} {\bibfnamefont {S.~K.}\ \bibnamefont
  {Lee}}, \bibinfo {author} {\bibfnamefont {M.}~\bibnamefont {Lisanti}},
  \bibinfo {author} {\bibfnamefont {S.}~\bibnamefont {Mishra-Sharma}}, \ and\
  \bibinfo {author} {\bibfnamefont {B.~R.}\ \bibnamefont {Safdi}},\ }\href
  {\doibase 10.1103/PhysRevD.92.083517} {\bibfield  {journal} {\bibinfo
  {journal} {Phys. Rev. D}\ }\textbf {\bibinfo {volume} {92}},\ \bibinfo
  {pages} {083517} (\bibinfo {year} {2015})},\ \Eprint
  {http://arxiv.org/abs/1508.07361} {arXiv:1508.07361 [hep-ph]} \BibitemShut
  {NoStop}%
\bibitem [{\citenamefont {Essig}\ \emph {et~al.}(2017)\citenamefont {Essig},
  \citenamefont {Volansky},\ and\ \citenamefont {Yu}}]{Essig:2017kqs}%
  \BibitemOpen
  \bibfield  {author} {\bibinfo {author} {\bibfnamefont {R.}~\bibnamefont
  {Essig}}, \bibinfo {author} {\bibfnamefont {T.}~\bibnamefont {Volansky}}, \
  and\ \bibinfo {author} {\bibfnamefont {T.-T.}\ \bibnamefont {Yu}},\ }\href
  {\doibase 10.1103/physrevd.96.043017} {\bibfield  {journal} {\bibinfo
  {journal} {Physical Review D}\ }\textbf {\bibinfo {volume} {96}},\ \bibinfo
  {pages} {043017} (\bibinfo {year} {2017})},\ \Eprint
  {http://arxiv.org/abs/1703.00910} {arXiv:1703.00910 [hep-ph]} \BibitemShut
  {NoStop}%
\bibitem [{\citenamefont {Catena}\ \emph {et~al.}(2020)\citenamefont {Catena},
  \citenamefont {Emken}, \citenamefont {Spaldin},\ and\ \citenamefont
  {Tarantino}}]{Catena:2019gfa}%
  \BibitemOpen
  \bibfield  {author} {\bibinfo {author} {\bibfnamefont {R.}~\bibnamefont
  {Catena}}, \bibinfo {author} {\bibfnamefont {T.}~\bibnamefont {Emken}},
  \bibinfo {author} {\bibfnamefont {N.~A.}\ \bibnamefont {Spaldin}}, \ and\
  \bibinfo {author} {\bibfnamefont {W.}~\bibnamefont {Tarantino}},\ }\href
  {\doibase 10.1103/PhysRevResearch.2.033195} {\bibfield  {journal} {\bibinfo
  {journal} {Phys. Rev. Res.}\ }\textbf {\bibinfo {volume} {2}},\ \bibinfo
  {pages} {033195} (\bibinfo {year} {2020})},\ \Eprint
  {http://arxiv.org/abs/1912.08204} {arXiv:1912.08204 [hep-ph]} \BibitemShut
  {NoStop}%
\bibitem [{\citenamefont {Agnes}\ \emph {et~al.}(2018)\citenamefont {Agnes}
  \emph {et~al.}}]{Agnes:2018oej}%
  \BibitemOpen
  \bibfield  {author} {\bibinfo {author} {\bibfnamefont {P.}~\bibnamefont
  {Agnes}} \emph {et~al.} (\bibinfo {collaboration} {DarkSide}),\ }\href
  {\doibase 10.1103/PhysRevLett.121.111303} {\bibfield  {journal} {\bibinfo
  {journal} {Phys. Rev. Lett.}\ }\textbf {\bibinfo {volume} {121}},\ \bibinfo
  {pages} {111303} (\bibinfo {year} {2018})},\ \Eprint
  {http://arxiv.org/abs/1802.06998} {arXiv:1802.06998 [astro-ph.CO]}
  \BibitemShut {NoStop}%
\bibitem [{\citenamefont {Aprile}\ \emph {et~al.}(2019)\citenamefont {Aprile}
  \emph {et~al.}}]{Aprile:2019xxb}%
  \BibitemOpen
  \bibfield  {author} {\bibinfo {author} {\bibfnamefont {E.}~\bibnamefont
  {Aprile}} \emph {et~al.} (\bibinfo {collaboration} {XENON}),\ }\href
  {\doibase 10.1103/PhysRevLett.123.251801} {\bibfield  {journal} {\bibinfo
  {journal} {Phys. Rev. Lett.}\ }\textbf {\bibinfo {volume} {123}},\ \bibinfo
  {pages} {251801} (\bibinfo {year} {2019})},\ \Eprint
  {http://arxiv.org/abs/1907.11485} {arXiv:1907.11485 [hep-ex]} \BibitemShut
  {NoStop}%
\bibitem [{\citenamefont {Aprile}\ \emph {et~al.}(2020)\citenamefont {Aprile}
  \emph {et~al.}}]{Aprile:2020tmw}%
  \BibitemOpen
  \bibfield  {author} {\bibinfo {author} {\bibfnamefont {E.}~\bibnamefont
  {Aprile}} \emph {et~al.} (\bibinfo {collaboration} {XENON}),\ }\href
  {\doibase 10.1103/PhysRevD.102.072004} {\bibfield  {journal} {\bibinfo
  {journal} {Phys. Rev. D}\ }\textbf {\bibinfo {volume} {102}},\ \bibinfo
  {pages} {072004} (\bibinfo {year} {2020})},\ \Eprint
  {http://arxiv.org/abs/2006.09721} {arXiv:2006.09721 [hep-ex]} \BibitemShut
  {NoStop}%
\bibitem [{\citenamefont {Essig}\ \emph
  {et~al.}(2012{\natexlab{b}})\citenamefont {Essig}, \citenamefont
  {Manalaysay}, \citenamefont {Mardon}, \citenamefont {Sorensen},\ and\
  \citenamefont {Volansky}}]{Essig:2012yx}%
  \BibitemOpen
  \bibfield  {author} {\bibinfo {author} {\bibfnamefont {R.}~\bibnamefont
  {Essig}}, \bibinfo {author} {\bibfnamefont {A.}~\bibnamefont {Manalaysay}},
  \bibinfo {author} {\bibfnamefont {J.}~\bibnamefont {Mardon}}, \bibinfo
  {author} {\bibfnamefont {P.}~\bibnamefont {Sorensen}}, \ and\ \bibinfo
  {author} {\bibfnamefont {T.}~\bibnamefont {Volansky}},\ }\href {\doibase
  10.1103/PhysRevLett.109.021301} {\bibfield  {journal} {\bibinfo  {journal}
  {Phys. Rev. Lett.}\ }\textbf {\bibinfo {volume} {109}},\ \bibinfo {pages}
  {021301} (\bibinfo {year} {2012}{\natexlab{b}})},\ \Eprint
  {http://arxiv.org/abs/1206.2644} {arXiv:1206.2644 [astro-ph.CO]} \BibitemShut
  {NoStop}%
\bibitem [{\citenamefont {Essig}\ \emph {et~al.}(2016)\citenamefont {Essig},
  \citenamefont {Fernandez-Serra}, \citenamefont {Mardon}, \citenamefont
  {Soto}, \citenamefont {Volansky},\ and\ \citenamefont {Yu}}]{Essig:2015cda}%
  \BibitemOpen
  \bibfield  {author} {\bibinfo {author} {\bibfnamefont {R.}~\bibnamefont
  {Essig}}, \bibinfo {author} {\bibfnamefont {M.}~\bibnamefont
  {Fernandez-Serra}}, \bibinfo {author} {\bibfnamefont {J.}~\bibnamefont
  {Mardon}}, \bibinfo {author} {\bibfnamefont {A.}~\bibnamefont {Soto}},
  \bibinfo {author} {\bibfnamefont {T.}~\bibnamefont {Volansky}}, \ and\
  \bibinfo {author} {\bibfnamefont {T.-T.}\ \bibnamefont {Yu}},\ }\href
  {\doibase 10.1007/JHEP05(2016)046} {\bibfield  {journal} {\bibinfo  {journal}
  {JHEP}\ }\textbf {\bibinfo {volume} {05}},\ \bibinfo {pages} {046} (\bibinfo
  {year} {2016})},\ \Eprint {http://arxiv.org/abs/1509.01598} {arXiv:1509.01598
  [hep-ph]} \BibitemShut {NoStop}%
\bibitem [{\citenamefont {Derenzo}\ \emph {et~al.}(2017)\citenamefont
  {Derenzo}, \citenamefont {Essig}, \citenamefont {Massari}, \citenamefont
  {Soto},\ and\ \citenamefont {Yu}}]{Derenzo:2016fse}%
  \BibitemOpen
  \bibfield  {author} {\bibinfo {author} {\bibfnamefont {S.}~\bibnamefont
  {Derenzo}}, \bibinfo {author} {\bibfnamefont {R.}~\bibnamefont {Essig}},
  \bibinfo {author} {\bibfnamefont {A.}~\bibnamefont {Massari}}, \bibinfo
  {author} {\bibfnamefont {A.}~\bibnamefont {Soto}}, \ and\ \bibinfo {author}
  {\bibfnamefont {T.-T.}\ \bibnamefont {Yu}},\ }\href {\doibase
  10.1103/PhysRevD.96.016026} {\bibfield  {journal} {\bibinfo  {journal} {Phys.
  Rev. D}\ }\textbf {\bibinfo {volume} {96}},\ \bibinfo {pages} {016026}
  (\bibinfo {year} {2017})},\ \Eprint {http://arxiv.org/abs/1607.01009}
  {arXiv:1607.01009 [hep-ph]} \BibitemShut {NoStop}%
\bibitem [{\citenamefont {Hochberg}\ \emph
  {et~al.}(2017{\natexlab{a}})\citenamefont {Hochberg}, \citenamefont {Lin},\
  and\ \citenamefont {Zurek}}]{Hochberg:2016sqx}%
  \BibitemOpen
  \bibfield  {author} {\bibinfo {author} {\bibfnamefont {Y.}~\bibnamefont
  {Hochberg}}, \bibinfo {author} {\bibfnamefont {T.}~\bibnamefont {Lin}}, \
  and\ \bibinfo {author} {\bibfnamefont {K.~M.}\ \bibnamefont {Zurek}},\ }\href
  {\doibase 10.1103/PhysRevD.95.023013} {\bibfield  {journal} {\bibinfo
  {journal} {Phys. Rev.}\ }\textbf {\bibinfo {volume} {D95}},\ \bibinfo {pages}
  {023013} (\bibinfo {year} {2017}{\natexlab{a}})},\ \Eprint
  {http://arxiv.org/abs/1608.01994} {arXiv:1608.01994 [hep-ph]} \BibitemShut
  {NoStop}%
\bibitem [{\citenamefont {Bloch}\ \emph {et~al.}(2017)\citenamefont {Bloch},
  \citenamefont {Essig}, \citenamefont {Tobioka}, \citenamefont {Volansky},\
  and\ \citenamefont {Yu}}]{Bloch:2016sjj}%
  \BibitemOpen
  \bibfield  {author} {\bibinfo {author} {\bibfnamefont {I.~M.}\ \bibnamefont
  {Bloch}}, \bibinfo {author} {\bibfnamefont {R.}~\bibnamefont {Essig}},
  \bibinfo {author} {\bibfnamefont {K.}~\bibnamefont {Tobioka}}, \bibinfo
  {author} {\bibfnamefont {T.}~\bibnamefont {Volansky}}, \ and\ \bibinfo
  {author} {\bibfnamefont {T.-T.}\ \bibnamefont {Yu}},\ }\href {\doibase
  10.1007/JHEP06(2017)087} {\bibfield  {journal} {\bibinfo  {journal} {Journal
  of High Energy Physics}\ }\textbf {\bibinfo {volume} {2017}},\ \bibinfo
  {pages} {87} (\bibinfo {year} {2017})},\ \Eprint
  {http://arxiv.org/abs/1608.02123} {arXiv:1608.02123 [hep-ph]} \BibitemShut
  {NoStop}%
\bibitem [{\citenamefont {Kurinsky}\ \emph {et~al.}(2019)\citenamefont
  {Kurinsky}, \citenamefont {Yu}, \citenamefont {Hochberg},\ and\ \citenamefont
  {Cabrera}}]{Kurinsky:2019pgb}%
  \BibitemOpen
  \bibfield  {author} {\bibinfo {author} {\bibfnamefont {N.~A.}\ \bibnamefont
  {Kurinsky}}, \bibinfo {author} {\bibfnamefont {T.~C.}\ \bibnamefont {Yu}},
  \bibinfo {author} {\bibfnamefont {Y.}~\bibnamefont {Hochberg}}, \ and\
  \bibinfo {author} {\bibfnamefont {B.}~\bibnamefont {Cabrera}},\ }\href@noop
  {} {\  (\bibinfo {year} {2019})},\ \Eprint {http://arxiv.org/abs/1901.07569}
  {arXiv:1901.07569 [hep-ex]} \BibitemShut {NoStop}%
\bibitem [{\citenamefont {Trickle}\ \emph {et~al.}(2020)\citenamefont
  {Trickle}, \citenamefont {Zhang}, \citenamefont {Zurek}, \citenamefont
  {Inzani},\ and\ \citenamefont {Griffin}}]{Trickle:2019nya}%
  \BibitemOpen
  \bibfield  {author} {\bibinfo {author} {\bibfnamefont {T.}~\bibnamefont
  {Trickle}}, \bibinfo {author} {\bibfnamefont {Z.}~\bibnamefont {Zhang}},
  \bibinfo {author} {\bibfnamefont {K.~M.}\ \bibnamefont {Zurek}}, \bibinfo
  {author} {\bibfnamefont {K.}~\bibnamefont {Inzani}}, \ and\ \bibinfo {author}
  {\bibfnamefont {S.}~\bibnamefont {Griffin}},\ }\href {\doibase
  10.1007/JHEP03(2020)036} {\bibfield  {journal} {\bibinfo  {journal} {JHEP}\
  }\textbf {\bibinfo {volume} {03}},\ \bibinfo {pages} {036} (\bibinfo {year}
  {2020})},\ \Eprint {http://arxiv.org/abs/1910.08092} {arXiv:1910.08092
  [hep-ph]} \BibitemShut {NoStop}%
\bibitem [{\citenamefont {Griffin}\ \emph {et~al.}(2020)\citenamefont
  {Griffin}, \citenamefont {Inzani}, \citenamefont {Trickle}, \citenamefont
  {Zhang},\ and\ \citenamefont {Zurek}}]{Griffin:2019mvc}%
  \BibitemOpen
  \bibfield  {author} {\bibinfo {author} {\bibfnamefont {S.~M.}\ \bibnamefont
  {Griffin}}, \bibinfo {author} {\bibfnamefont {K.}~\bibnamefont {Inzani}},
  \bibinfo {author} {\bibfnamefont {T.}~\bibnamefont {Trickle}}, \bibinfo
  {author} {\bibfnamefont {Z.}~\bibnamefont {Zhang}}, \ and\ \bibinfo {author}
  {\bibfnamefont {K.~M.}\ \bibnamefont {Zurek}},\ }\href {\doibase
  10.1103/PhysRevD.101.055004} {\bibfield  {journal} {\bibinfo  {journal}
  {Phys. Rev. D}\ }\textbf {\bibinfo {volume} {101}},\ \bibinfo {pages}
  {055004} (\bibinfo {year} {2020})},\ \Eprint
  {http://arxiv.org/abs/1910.10716} {arXiv:1910.10716 [hep-ph]} \BibitemShut
  {NoStop}%
\bibitem [{\citenamefont {Griffin}\ \emph
  {et~al.}(2021{\natexlab{a}})\citenamefont {Griffin}, \citenamefont
  {Hochberg}, \citenamefont {Inzani}, \citenamefont {Kurinsky}, \citenamefont
  {Lin},\ and\ \citenamefont {Chin}}]{Griffin:2020lgd}%
  \BibitemOpen
  \bibfield  {author} {\bibinfo {author} {\bibfnamefont {S.~M.}\ \bibnamefont
  {Griffin}}, \bibinfo {author} {\bibfnamefont {Y.}~\bibnamefont {Hochberg}},
  \bibinfo {author} {\bibfnamefont {K.}~\bibnamefont {Inzani}}, \bibinfo
  {author} {\bibfnamefont {N.}~\bibnamefont {Kurinsky}}, \bibinfo {author}
  {\bibfnamefont {T.}~\bibnamefont {Lin}}, \ and\ \bibinfo {author}
  {\bibfnamefont {T.}~\bibnamefont {Chin}},\ }\href {\doibase
  10.1103/PhysRevD.103.075002} {\bibfield  {journal} {\bibinfo  {journal}
  {Phys. Rev. D}\ }\textbf {\bibinfo {volume} {103}},\ \bibinfo {pages}
  {075002} (\bibinfo {year} {2021}{\natexlab{a}})},\ \Eprint
  {http://arxiv.org/abs/2008.08560} {arXiv:2008.08560 [hep-ph]} \BibitemShut
  {NoStop}%
\bibitem [{\citenamefont {Du}\ \emph {et~al.}(2020)\citenamefont {Du},
  \citenamefont {Egana-Ugrinovic}, \citenamefont {Essig},\ and\ \citenamefont
  {Sholapurkar}}]{Du:2020ldo}%
  \BibitemOpen
  \bibfield  {author} {\bibinfo {author} {\bibfnamefont {P.}~\bibnamefont
  {Du}}, \bibinfo {author} {\bibfnamefont {D.}~\bibnamefont {Egana-Ugrinovic}},
  \bibinfo {author} {\bibfnamefont {R.}~\bibnamefont {Essig}}, \ and\ \bibinfo
  {author} {\bibfnamefont {M.}~\bibnamefont {Sholapurkar}},\ }\href@noop {} {\
  (\bibinfo {year} {2020})},\ \Eprint {http://arxiv.org/abs/2011.13939}
  {arXiv:2011.13939 [hep-ph]} \BibitemShut {NoStop}%
\bibitem [{\citenamefont {Mitridate}\ \emph {et~al.}(2021)\citenamefont
  {Mitridate}, \citenamefont {Trickle}, \citenamefont {Zhang},\ and\
  \citenamefont {Zurek}}]{Mitridate:2021ctr}%
  \BibitemOpen
  \bibfield  {author} {\bibinfo {author} {\bibfnamefont {A.}~\bibnamefont
  {Mitridate}}, \bibinfo {author} {\bibfnamefont {T.}~\bibnamefont {Trickle}},
  \bibinfo {author} {\bibfnamefont {Z.}~\bibnamefont {Zhang}}, \ and\ \bibinfo
  {author} {\bibfnamefont {K.~M.}\ \bibnamefont {Zurek}},\ }\href {\doibase
  10.1007/JHEP09(2021)123} {\bibfield  {journal} {\bibinfo  {journal} {JHEP}\
  }\textbf {\bibinfo {volume} {09}},\ \bibinfo {pages} {123} (\bibinfo {year}
  {2021})},\ \bibinfo {note} {arXiv: 2106.12586},\ \Eprint
  {http://arxiv.org/abs/2106.12586} {arXiv:2106.12586 [hep-ph]} \BibitemShut
  {NoStop}%
\bibitem [{\citenamefont {Hochberg}\ \emph
  {et~al.}(2016{\natexlab{a}})\citenamefont {Hochberg}, \citenamefont {Zhao},\
  and\ \citenamefont {Zurek}}]{Hochberg:2015pha}%
  \BibitemOpen
  \bibfield  {author} {\bibinfo {author} {\bibfnamefont {Y.}~\bibnamefont
  {Hochberg}}, \bibinfo {author} {\bibfnamefont {Y.}~\bibnamefont {Zhao}}, \
  and\ \bibinfo {author} {\bibfnamefont {K.~M.}\ \bibnamefont {Zurek}},\ }\href
  {\doibase 10.1103/PhysRevLett.116.011301} {\bibfield  {journal} {\bibinfo
  {journal} {Phys. Rev. Lett.}\ }\textbf {\bibinfo {volume} {116}},\ \bibinfo
  {pages} {011301} (\bibinfo {year} {2016}{\natexlab{a}})},\ \Eprint
  {http://arxiv.org/abs/1504.07237} {arXiv:1504.07237 [hep-ph]} \BibitemShut
  {NoStop}%
\bibitem [{\citenamefont {Hochberg}\ \emph
  {et~al.}(2016{\natexlab{b}})\citenamefont {Hochberg}, \citenamefont {Lin},\
  and\ \citenamefont {Zurek}}]{Hochberg:2016ajh}%
  \BibitemOpen
  \bibfield  {author} {\bibinfo {author} {\bibfnamefont {Y.}~\bibnamefont
  {Hochberg}}, \bibinfo {author} {\bibfnamefont {T.}~\bibnamefont {Lin}}, \
  and\ \bibinfo {author} {\bibfnamefont {K.~M.}\ \bibnamefont {Zurek}},\ }\href
  {\doibase 10.1103/PhysRevD.94.015019} {\bibfield  {journal} {\bibinfo
  {journal} {Phys. Rev.}\ }\textbf {\bibinfo {volume} {D94}},\ \bibinfo {pages}
  {015019} (\bibinfo {year} {2016}{\natexlab{b}})},\ \Eprint
  {http://arxiv.org/abs/1604.06800} {arXiv:1604.06800 [hep-ph]} \BibitemShut
  {NoStop}%
\bibitem [{\citenamefont {Hochberg}\ \emph
  {et~al.}(2017{\natexlab{b}})\citenamefont {Hochberg}, \citenamefont {Kahn},
  \citenamefont {Lisanti}, \citenamefont {Tully},\ and\ \citenamefont
  {Zurek}}]{Hochberg:2016ntt}%
  \BibitemOpen
  \bibfield  {author} {\bibinfo {author} {\bibfnamefont {Y.}~\bibnamefont
  {Hochberg}}, \bibinfo {author} {\bibfnamefont {Y.}~\bibnamefont {Kahn}},
  \bibinfo {author} {\bibfnamefont {M.}~\bibnamefont {Lisanti}}, \bibinfo
  {author} {\bibfnamefont {C.~G.}\ \bibnamefont {Tully}}, \ and\ \bibinfo
  {author} {\bibfnamefont {K.~M.}\ \bibnamefont {Zurek}},\ }\href {\doibase
  10.1016/j.physletb.2017.06.051} {\bibfield  {journal} {\bibinfo  {journal}
  {Phys. Lett. B}\ }\textbf {\bibinfo {volume} {772}},\ \bibinfo {pages} {239}
  (\bibinfo {year} {2017}{\natexlab{b}})},\ \Eprint
  {http://arxiv.org/abs/1606.08849} {arXiv:1606.08849 [hep-ph]} \BibitemShut
  {NoStop}%
\bibitem [{\citenamefont {Hochberg}\ \emph
  {et~al.}(2016{\natexlab{c}})\citenamefont {Hochberg}, \citenamefont {Pyle},
  \citenamefont {Zhao},\ and\ \citenamefont {Zurek}}]{Hochberg:2015fth}%
  \BibitemOpen
  \bibfield  {author} {\bibinfo {author} {\bibfnamefont {Y.}~\bibnamefont
  {Hochberg}}, \bibinfo {author} {\bibfnamefont {M.}~\bibnamefont {Pyle}},
  \bibinfo {author} {\bibfnamefont {Y.}~\bibnamefont {Zhao}}, \ and\ \bibinfo
  {author} {\bibfnamefont {K.~M.}\ \bibnamefont {Zurek}},\ }\href {\doibase
  10.1007/JHEP08(2016)057} {\bibfield  {journal} {\bibinfo  {journal} {JHEP}\
  }\textbf {\bibinfo {volume} {08}},\ \bibinfo {pages} {057} (\bibinfo {year}
  {2016}{\natexlab{c}})},\ \Eprint {http://arxiv.org/abs/1512.04533}
  {arXiv:1512.04533 [hep-ph]} \BibitemShut {NoStop}%
\bibitem [{\citenamefont {Hochberg}\ \emph {et~al.}(2018)\citenamefont
  {Hochberg}, \citenamefont {Kahn}, \citenamefont {Lisanti}, \citenamefont
  {Zurek}, \citenamefont {Grushin}, \citenamefont {Ilan}, \citenamefont
  {Griffin}, \citenamefont {Liu}, \citenamefont {Weber},\ and\ \citenamefont
  {Neaton}}]{Hochberg:2017wce}%
  \BibitemOpen
  \bibfield  {author} {\bibinfo {author} {\bibfnamefont {Y.}~\bibnamefont
  {Hochberg}}, \bibinfo {author} {\bibfnamefont {Y.}~\bibnamefont {Kahn}},
  \bibinfo {author} {\bibfnamefont {M.}~\bibnamefont {Lisanti}}, \bibinfo
  {author} {\bibfnamefont {K.~M.}\ \bibnamefont {Zurek}}, \bibinfo {author}
  {\bibfnamefont {A.~G.}\ \bibnamefont {Grushin}}, \bibinfo {author}
  {\bibfnamefont {R.}~\bibnamefont {Ilan}}, \bibinfo {author} {\bibfnamefont
  {S.~M.}\ \bibnamefont {Griffin}}, \bibinfo {author} {\bibfnamefont {Z.-F.}\
  \bibnamefont {Liu}}, \bibinfo {author} {\bibfnamefont {S.~F.}\ \bibnamefont
  {Weber}}, \ and\ \bibinfo {author} {\bibfnamefont {J.~B.}\ \bibnamefont
  {Neaton}},\ }\href {\doibase 10.1103/PhysRevD.97.015004} {\bibfield
  {journal} {\bibinfo  {journal} {Physical Review D}\ }\textbf {\bibinfo
  {volume} {97}},\ \bibinfo {pages} {015004} (\bibinfo {year} {2018})},\
  \Eprint {http://arxiv.org/abs/1708.08929} {arXiv:1708.08929 [hep-ph]}
  \BibitemShut {NoStop}%
\bibitem [{\citenamefont {Coskuner}\ \emph {et~al.}(2019)\citenamefont
  {Coskuner}, \citenamefont {Mitridate}, \citenamefont {Olivares},\ and\
  \citenamefont {Zurek}}]{Coskuner:2019odd}%
  \BibitemOpen
  \bibfield  {author} {\bibinfo {author} {\bibfnamefont {A.}~\bibnamefont
  {Coskuner}}, \bibinfo {author} {\bibfnamefont {A.}~\bibnamefont {Mitridate}},
  \bibinfo {author} {\bibfnamefont {A.}~\bibnamefont {Olivares}}, \ and\
  \bibinfo {author} {\bibfnamefont {K.~M.}\ \bibnamefont {Zurek}},\ }\href
  {\doibase 10.1103/PhysRevD.103.016006} {\bibfield  {journal} {\bibinfo
  {journal} {Phys. Rev. D}\ }\textbf {\bibinfo {volume} {103}},\ \bibinfo
  {pages} {016006} (\bibinfo {year} {2019})},\ \Eprint
  {http://arxiv.org/abs/1909.09170} {arXiv:1909.09170 [hep-ph]} \BibitemShut
  {NoStop}%
\bibitem [{\citenamefont {Geilhufe}\ \emph {et~al.}(2019)\citenamefont
  {Geilhufe}, \citenamefont {Kahlhoefer},\ and\ \citenamefont
  {Winkler}}]{Geilhufe:2019ndy}%
  \BibitemOpen
  \bibfield  {author} {\bibinfo {author} {\bibfnamefont {R.~M.}\ \bibnamefont
  {Geilhufe}}, \bibinfo {author} {\bibfnamefont {F.}~\bibnamefont
  {Kahlhoefer}}, \ and\ \bibinfo {author} {\bibfnamefont {M.~W.}\ \bibnamefont
  {Winkler}},\ }\href {\doibase 10.1103/PhysRevD.101.055005} {\bibfield
  {journal} {\bibinfo  {journal} {Phys. Rev. D}\ }\textbf {\bibinfo {volume}
  {101}},\ \bibinfo {pages} {055005} (\bibinfo {year} {2019})},\ \Eprint
  {http://arxiv.org/abs/1910.02091} {arXiv:1910.02091 [hep-ph]} \BibitemShut
  {NoStop}%
\bibitem [{\citenamefont {Inzani}\ \emph {et~al.}(2021)\citenamefont {Inzani},
  \citenamefont {Faghaninia},\ and\ \citenamefont {Griffin}}]{Inzani:2020szg}%
  \BibitemOpen
  \bibfield  {author} {\bibinfo {author} {\bibfnamefont {K.}~\bibnamefont
  {Inzani}}, \bibinfo {author} {\bibfnamefont {A.}~\bibnamefont {Faghaninia}},
  \ and\ \bibinfo {author} {\bibfnamefont {S.~M.}\ \bibnamefont {Griffin}},\
  }\href {\doibase 10.1103/PhysRevResearch.3.013069} {\bibfield  {journal}
  {\bibinfo  {journal} {Physical Review Research}\ }\textbf {\bibinfo {volume}
  {3}},\ \bibinfo {pages} {013069} (\bibinfo {year} {2021})},\ \Eprint
  {http://arxiv.org/abs/2008.05062} {arXiv:2008.05062 [cond-mat.mtrl-sci]}
  \BibitemShut {NoStop}%
\bibitem [{\citenamefont {Blanco}\ \emph {et~al.}(2020)\citenamefont {Blanco},
  \citenamefont {Collar}, \citenamefont {Kahn},\ and\ \citenamefont
  {Lillard}}]{Blanco:2019lrf}%
  \BibitemOpen
  \bibfield  {author} {\bibinfo {author} {\bibfnamefont {C.}~\bibnamefont
  {Blanco}}, \bibinfo {author} {\bibfnamefont {J.~I.}\ \bibnamefont {Collar}},
  \bibinfo {author} {\bibfnamefont {Y.}~\bibnamefont {Kahn}}, \ and\ \bibinfo
  {author} {\bibfnamefont {B.}~\bibnamefont {Lillard}},\ }\href {\doibase
  10.1103/PhysRevD.101.056001} {\bibfield  {journal} {\bibinfo  {journal}
  {Phys. Rev. D}\ }\textbf {\bibinfo {volume} {101}},\ \bibinfo {pages}
  {056001} (\bibinfo {year} {2020})},\ \Eprint
  {http://arxiv.org/abs/1912.02822} {arXiv:1912.02822 [hep-ph]} \BibitemShut
  {NoStop}%
\bibitem [{\citenamefont {Blanco}\ \emph {et~al.}(2021)\citenamefont {Blanco},
  \citenamefont {Kahn}, \citenamefont {Lillard},\ and\ \citenamefont
  {McDermott}}]{Blanco:2021hlm}%
  \BibitemOpen
  \bibfield  {author} {\bibinfo {author} {\bibfnamefont {C.}~\bibnamefont
  {Blanco}}, \bibinfo {author} {\bibfnamefont {Y.}~\bibnamefont {Kahn}},
  \bibinfo {author} {\bibfnamefont {B.}~\bibnamefont {Lillard}}, \ and\
  \bibinfo {author} {\bibfnamefont {S.~D.}\ \bibnamefont {McDermott}},\ }\href
  {\doibase 10.1103/PhysRevD.104.036011} {\bibfield  {journal} {\bibinfo
  {journal} {Phys. Rev. D}\ }\textbf {\bibinfo {volume} {104}},\ \bibinfo
  {pages} {036011} (\bibinfo {year} {2021})},\ \Eprint
  {http://arxiv.org/abs/2103.08601} {arXiv:2103.08601 [hep-ph]} \BibitemShut
  {NoStop}%
\bibitem [{\citenamefont {{EXCEED-DM Collaboration}}(2021)}]{EDC2021}%
  \BibitemOpen
  \bibfield  {author} {\bibinfo {author} {\bibnamefont {{EXCEED-DM
  Collaboration}}},\ }\href {\doibase 10.5281/ZENODO.4747695} {\enquote
  {\bibinfo {title} {tanner-trickle/exceed-dm: Exceed-dm-v0.3.0},}\ } (\bibinfo
  {year} {2021})\BibitemShut {NoStop}%
\bibitem [{\citenamefont {Griffin}\ \emph
  {et~al.}(2021{\natexlab{b}})\citenamefont {Griffin}, \citenamefont {Inzani},
  \citenamefont {Trickle}, \citenamefont {Zhang},\ and\ \citenamefont
  {Zurek}}]{Griffin:2021znd}%
  \BibitemOpen
  \bibfield  {author} {\bibinfo {author} {\bibfnamefont {S.~M.}\ \bibnamefont
  {Griffin}}, \bibinfo {author} {\bibfnamefont {K.}~\bibnamefont {Inzani}},
  \bibinfo {author} {\bibfnamefont {T.}~\bibnamefont {Trickle}}, \bibinfo
  {author} {\bibfnamefont {Z.}~\bibnamefont {Zhang}}, \ and\ \bibinfo {author}
  {\bibfnamefont {K.~M.}\ \bibnamefont {Zurek}},\ }\href {\doibase
  10.1103/PhysRevD.104.095015} {\bibfield  {journal} {\bibinfo  {journal}
  {Phys. Rev. D}\ }\textbf {\bibinfo {volume} {104}},\ \bibinfo {pages}
  {095015} (\bibinfo {year} {2021}{\natexlab{b}})},\ \Eprint
  {http://arxiv.org/abs/2105.05253} {arXiv:2105.05253 [hep-ph]} \BibitemShut
  {NoStop}%
\bibitem [{\citenamefont {Zheng}\ \emph {et~al.}(2016)\citenamefont {Zheng},
  \citenamefont {Lu}, \citenamefont {Zhu}, \citenamefont {Ning}, \citenamefont
  {Han}, \citenamefont {Zhang}, \citenamefont {Zhang}, \citenamefont {Xi},
  \citenamefont {Yang}, \citenamefont {Du}, \citenamefont {Yang}, \citenamefont
  {Zhang},\ and\ \citenamefont {Tian}}]{Zheng2016}%
  \BibitemOpen
  \bibfield  {author} {\bibinfo {author} {\bibfnamefont {G.}~\bibnamefont
  {Zheng}}, \bibinfo {author} {\bibfnamefont {J.}~\bibnamefont {Lu}}, \bibinfo
  {author} {\bibfnamefont {X.}~\bibnamefont {Zhu}}, \bibinfo {author}
  {\bibfnamefont {W.}~\bibnamefont {Ning}}, \bibinfo {author} {\bibfnamefont
  {Y.}~\bibnamefont {Han}}, \bibinfo {author} {\bibfnamefont {H.}~\bibnamefont
  {Zhang}}, \bibinfo {author} {\bibfnamefont {J.}~\bibnamefont {Zhang}},
  \bibinfo {author} {\bibfnamefont {C.}~\bibnamefont {Xi}}, \bibinfo {author}
  {\bibfnamefont {J.}~\bibnamefont {Yang}}, \bibinfo {author} {\bibfnamefont
  {H.}~\bibnamefont {Du}}, \bibinfo {author} {\bibfnamefont {K.}~\bibnamefont
  {Yang}}, \bibinfo {author} {\bibfnamefont {Y.}~\bibnamefont {Zhang}}, \ and\
  \bibinfo {author} {\bibfnamefont {M.}~\bibnamefont {Tian}},\ }\href {\doibase
  10.1103/PhysRevB.93.115414} {\bibfield  {journal} {\bibinfo  {journal}
  {Physical Review B}\ }\textbf {\bibinfo {volume} {93}},\ \bibinfo {pages}
  {115414} (\bibinfo {year} {2016})}\BibitemShut {NoStop}%
\bibitem [{\citenamefont {Chen}\ \emph
  {et~al.}(2015{\natexlab{a}})\citenamefont {Chen}, \citenamefont {Chen},
  \citenamefont {Song}, \citenamefont {Schneeloch}, \citenamefont {Gu},
  \citenamefont {Wang},\ and\ \citenamefont {Wang}}]{Chen2015}%
  \BibitemOpen
  \bibfield  {author} {\bibinfo {author} {\bibfnamefont {R.~Y.}\ \bibnamefont
  {Chen}}, \bibinfo {author} {\bibfnamefont {Z.~G.}\ \bibnamefont {Chen}},
  \bibinfo {author} {\bibfnamefont {X.-Y.}\ \bibnamefont {Song}}, \bibinfo
  {author} {\bibfnamefont {J.~A.}\ \bibnamefont {Schneeloch}}, \bibinfo
  {author} {\bibfnamefont {G.~D.}\ \bibnamefont {Gu}}, \bibinfo {author}
  {\bibfnamefont {F.}~\bibnamefont {Wang}}, \ and\ \bibinfo {author}
  {\bibfnamefont {N.~L.}\ \bibnamefont {Wang}},\ }\href {\doibase
  10.1103/PhysRevLett.115.176404} {\bibfield  {journal} {\bibinfo  {journal}
  {Physical Review Letters}\ }\textbf {\bibinfo {volume} {115}},\ \bibinfo
  {pages} {176404} (\bibinfo {year} {2015}{\natexlab{a}})}\BibitemShut
  {NoStop}%
\bibitem [{\citenamefont {Liu}\ \emph {et~al.}(2018)\citenamefont {Liu},
  \citenamefont {Wang}, \citenamefont {Chen}, \citenamefont {Xu}, \citenamefont
  {Jiang}, \citenamefont {Yang}, \citenamefont {Yang}, \citenamefont {Lv},
  \citenamefont {Zhou}, \citenamefont {Chen}, \citenamefont {Yao},
  \citenamefont {Lu}, \citenamefont {Chen}, \citenamefont {Felser},
  \citenamefont {Yan}, \citenamefont {Liu},\ and\ \citenamefont
  {Chen}}]{Liu2018}%
  \BibitemOpen
  \bibfield  {author} {\bibinfo {author} {\bibfnamefont {S.}~\bibnamefont
  {Liu}}, \bibinfo {author} {\bibfnamefont {M.~X.}\ \bibnamefont {Wang}},
  \bibinfo {author} {\bibfnamefont {C.}~\bibnamefont {Chen}}, \bibinfo {author}
  {\bibfnamefont {X.}~\bibnamefont {Xu}}, \bibinfo {author} {\bibfnamefont
  {J.}~\bibnamefont {Jiang}}, \bibinfo {author} {\bibfnamefont {L.~X.}\
  \bibnamefont {Yang}}, \bibinfo {author} {\bibfnamefont {H.~F.}\ \bibnamefont
  {Yang}}, \bibinfo {author} {\bibfnamefont {Y.~Y.}\ \bibnamefont {Lv}},
  \bibinfo {author} {\bibfnamefont {J.}~\bibnamefont {Zhou}}, \bibinfo {author}
  {\bibfnamefont {Y.~B.}\ \bibnamefont {Chen}}, \bibinfo {author}
  {\bibfnamefont {S.~H.}\ \bibnamefont {Yao}}, \bibinfo {author} {\bibfnamefont
  {M.~H.}\ \bibnamefont {Lu}}, \bibinfo {author} {\bibfnamefont {Y.~F.}\
  \bibnamefont {Chen}}, \bibinfo {author} {\bibfnamefont {C.}~\bibnamefont
  {Felser}}, \bibinfo {author} {\bibfnamefont {B.~H.}\ \bibnamefont {Yan}},
  \bibinfo {author} {\bibfnamefont {Z.~K.}\ \bibnamefont {Liu}}, \ and\
  \bibinfo {author} {\bibfnamefont {Y.~L.}\ \bibnamefont {Chen}},\ }\href
  {\doibase 10.1063/1.5050847} {\bibfield  {journal} {\bibinfo  {journal}
  {{APL} Materials}\ }\textbf {\bibinfo {volume} {6}},\ \bibinfo {pages}
  {121111} (\bibinfo {year} {2018})}\BibitemShut {NoStop}%
\bibitem [{\citenamefont {Li}\ \emph {et~al.}(2016)\citenamefont {Li},
  \citenamefont {Kharzeev}, \citenamefont {Zhang}, \citenamefont {Huang},
  \citenamefont {Pletikosic}, \citenamefont {Fedorov}, \citenamefont {Zhong},
  \citenamefont {Schneeloch}, \citenamefont {Gu},\ and\ \citenamefont
  {Valla}}]{Li:2014bha}%
  \BibitemOpen
  \bibfield  {author} {\bibinfo {author} {\bibfnamefont {Q.}~\bibnamefont
  {Li}}, \bibinfo {author} {\bibfnamefont {D.~E.}\ \bibnamefont {Kharzeev}},
  \bibinfo {author} {\bibfnamefont {C.}~\bibnamefont {Zhang}}, \bibinfo
  {author} {\bibfnamefont {Y.}~\bibnamefont {Huang}}, \bibinfo {author}
  {\bibfnamefont {I.}~\bibnamefont {Pletikosic}}, \bibinfo {author}
  {\bibfnamefont {A.~V.}\ \bibnamefont {Fedorov}}, \bibinfo {author}
  {\bibfnamefont {R.~D.}\ \bibnamefont {Zhong}}, \bibinfo {author}
  {\bibfnamefont {J.~A.}\ \bibnamefont {Schneeloch}}, \bibinfo {author}
  {\bibfnamefont {G.~D.}\ \bibnamefont {Gu}}, \ and\ \bibinfo {author}
  {\bibfnamefont {T.}~\bibnamefont {Valla}},\ }\href {\doibase
  10.1038/nphys3648} {\bibfield  {journal} {\bibinfo  {journal} {Nature Phys.}\
  }\textbf {\bibinfo {volume} {12}},\ \bibinfo {pages} {550} (\bibinfo {year}
  {2016})},\ \Eprint {http://arxiv.org/abs/1412.6543} {arXiv:1412.6543
  [cond-mat.str-el]} \BibitemShut {NoStop}%
\bibitem [{\citenamefont {Chen}\ \emph
  {et~al.}(2015{\natexlab{b}})\citenamefont {Chen}, \citenamefont {Zhang},
  \citenamefont {Schneeloch}, \citenamefont {Zhang}, \citenamefont {Li},
  \citenamefont {Gu},\ and\ \citenamefont {Wang}}]{Chen2015a}%
  \BibitemOpen
  \bibfield  {author} {\bibinfo {author} {\bibfnamefont {R.~Y.}\ \bibnamefont
  {Chen}}, \bibinfo {author} {\bibfnamefont {S.~J.}\ \bibnamefont {Zhang}},
  \bibinfo {author} {\bibfnamefont {J.~A.}\ \bibnamefont {Schneeloch}},
  \bibinfo {author} {\bibfnamefont {C.}~\bibnamefont {Zhang}}, \bibinfo
  {author} {\bibfnamefont {Q.}~\bibnamefont {Li}}, \bibinfo {author}
  {\bibfnamefont {G.~D.}\ \bibnamefont {Gu}}, \ and\ \bibinfo {author}
  {\bibfnamefont {N.~L.}\ \bibnamefont {Wang}},\ }\href {\doibase
  10.1103/PhysRevB.92.075107} {\bibfield  {journal} {\bibinfo  {journal}
  {Physical Review B}\ }\textbf {\bibinfo {volume} {92}},\ \bibinfo {pages}
  {075107} (\bibinfo {year} {2015}{\natexlab{b}})}\BibitemShut {NoStop}%
\bibitem [{\citenamefont {Yuan}\ \emph {et~al.}(2016)\citenamefont {Yuan},
  \citenamefont {Zhang}, \citenamefont {Liu}, \citenamefont {Narayan},
  \citenamefont {Song}, \citenamefont {Shen}, \citenamefont {Sui},
  \citenamefont {Xu}, \citenamefont {Yu}, \citenamefont {An}, \citenamefont
  {Zhao}, \citenamefont {Sanvito}, \citenamefont {Yan},\ and\ \citenamefont
  {Xiu}}]{Yuan2016}%
  \BibitemOpen
  \bibfield  {author} {\bibinfo {author} {\bibfnamefont {X.}~\bibnamefont
  {Yuan}}, \bibinfo {author} {\bibfnamefont {C.}~\bibnamefont {Zhang}},
  \bibinfo {author} {\bibfnamefont {Y.}~\bibnamefont {Liu}}, \bibinfo {author}
  {\bibfnamefont {A.}~\bibnamefont {Narayan}}, \bibinfo {author} {\bibfnamefont
  {C.}~\bibnamefont {Song}}, \bibinfo {author} {\bibfnamefont {S.}~\bibnamefont
  {Shen}}, \bibinfo {author} {\bibfnamefont {X.}~\bibnamefont {Sui}}, \bibinfo
  {author} {\bibfnamefont {J.}~\bibnamefont {Xu}}, \bibinfo {author}
  {\bibfnamefont {H.}~\bibnamefont {Yu}}, \bibinfo {author} {\bibfnamefont
  {Z.}~\bibnamefont {An}}, \bibinfo {author} {\bibfnamefont {J.}~\bibnamefont
  {Zhao}}, \bibinfo {author} {\bibfnamefont {S.}~\bibnamefont {Sanvito}},
  \bibinfo {author} {\bibfnamefont {H.}~\bibnamefont {Yan}}, \ and\ \bibinfo
  {author} {\bibfnamefont {F.}~\bibnamefont {Xiu}},\ }\href {\doibase
  10.1038/am.2016.166} {\bibfield  {journal} {\bibinfo  {journal} {NPG Asia
  Materials}\ }\textbf {\bibinfo {volume} {8}},\ \bibinfo {pages} {e325}
  (\bibinfo {year} {2016})}\BibitemShut {NoStop}%
\bibitem [{\citenamefont {Chen}\ \emph {et~al.}(2017)\citenamefont {Chen},
  \citenamefont {Chen}, \citenamefont {Zhong}, \citenamefont {Schneeloch},
  \citenamefont {Zhang}, \citenamefont {Huang}, \citenamefont {Qu},
  \citenamefont {Yu}, \citenamefont {Li}, \citenamefont {Gu},\ and\
  \citenamefont {Wang}}]{Chen2017}%
  \BibitemOpen
  \bibfield  {author} {\bibinfo {author} {\bibfnamefont {Z.-G.}\ \bibnamefont
  {Chen}}, \bibinfo {author} {\bibfnamefont {R.~Y.}\ \bibnamefont {Chen}},
  \bibinfo {author} {\bibfnamefont {R.~D.}\ \bibnamefont {Zhong}}, \bibinfo
  {author} {\bibfnamefont {J.}~\bibnamefont {Schneeloch}}, \bibinfo {author}
  {\bibfnamefont {C.}~\bibnamefont {Zhang}}, \bibinfo {author} {\bibfnamefont
  {Y.}~\bibnamefont {Huang}}, \bibinfo {author} {\bibfnamefont
  {F.}~\bibnamefont {Qu}}, \bibinfo {author} {\bibfnamefont {R.}~\bibnamefont
  {Yu}}, \bibinfo {author} {\bibfnamefont {Q.}~\bibnamefont {Li}}, \bibinfo
  {author} {\bibfnamefont {G.~D.}\ \bibnamefont {Gu}}, \ and\ \bibinfo {author}
  {\bibfnamefont {N.~L.}\ \bibnamefont {Wang}},\ }\href {\doibase
  10.1073/pnas.1613110114} {\bibfield  {journal} {\bibinfo  {journal}
  {Proceedings of the National Academy of Sciences}\ }\textbf {\bibinfo
  {volume} {114}},\ \bibinfo {pages} {816} (\bibinfo {year}
  {2017})}\BibitemShut {NoStop}%
\bibitem [{\citenamefont {Monserrat}\ and\ \citenamefont
  {Narayan}(2019)}]{Monserrat2019}%
  \BibitemOpen
  \bibfield  {author} {\bibinfo {author} {\bibfnamefont {B.}~\bibnamefont
  {Monserrat}}\ and\ \bibinfo {author} {\bibfnamefont {A.}~\bibnamefont
  {Narayan}},\ }\href {\doibase 10.1103/PhysRevResearch.1.033181} {\bibfield
  {journal} {\bibinfo  {journal} {Phys. Rev. Research 1, 033181 (2019)}\ }
  (\bibinfo {year} {2019}),\ 10.1103/PhysRevResearch.1.033181},\ \Eprint
  {http://arxiv.org/abs/1909.07613} {arXiv:1909.07613 [cond-mat.mtrl-sci]}
  \BibitemShut {NoStop}%
\bibitem [{\citenamefont {Wu}\ \emph {et~al.}(2016)\citenamefont {Wu},
  \citenamefont {Ma}, \citenamefont {Nie}, \citenamefont {Zhao}, \citenamefont
  {Huang}, \citenamefont {Yin}, \citenamefont {Fu}, \citenamefont {Richard},
  \citenamefont {Chen}, \citenamefont {Fang}, \citenamefont {Dai},
  \citenamefont {Weng}, \citenamefont {Qian}, \citenamefont {Ding},\ and\
  \citenamefont {Pan}}]{Wu2016}%
  \BibitemOpen
  \bibfield  {author} {\bibinfo {author} {\bibfnamefont {R.}~\bibnamefont
  {Wu}}, \bibinfo {author} {\bibfnamefont {J.-Z.}\ \bibnamefont {Ma}}, \bibinfo
  {author} {\bibfnamefont {S.-M.}\ \bibnamefont {Nie}}, \bibinfo {author}
  {\bibfnamefont {L.-X.}\ \bibnamefont {Zhao}}, \bibinfo {author}
  {\bibfnamefont {X.}~\bibnamefont {Huang}}, \bibinfo {author} {\bibfnamefont
  {J.-X.}\ \bibnamefont {Yin}}, \bibinfo {author} {\bibfnamefont {B.-B.}\
  \bibnamefont {Fu}}, \bibinfo {author} {\bibfnamefont {P.}~\bibnamefont
  {Richard}}, \bibinfo {author} {\bibfnamefont {G.-F.}\ \bibnamefont {Chen}},
  \bibinfo {author} {\bibfnamefont {Z.}~\bibnamefont {Fang}}, \bibinfo {author}
  {\bibfnamefont {X.}~\bibnamefont {Dai}}, \bibinfo {author} {\bibfnamefont
  {H.-M.}\ \bibnamefont {Weng}}, \bibinfo {author} {\bibfnamefont
  {T.}~\bibnamefont {Qian}}, \bibinfo {author} {\bibfnamefont {H.}~\bibnamefont
  {Ding}}, \ and\ \bibinfo {author} {\bibfnamefont {S.~H.}\ \bibnamefont
  {Pan}},\ }\href {\doibase 10.1103/PhysRevX.6.021017} {\bibfield  {journal}
  {\bibinfo  {journal} {Physical Review X}\ }\textbf {\bibinfo {volume} {6}},\
  \bibinfo {pages} {021017} (\bibinfo {year} {2016})}\BibitemShut {NoStop}%
\bibitem [{\citenamefont {Nair}\ \emph {et~al.}(2017)\citenamefont {Nair},
  \citenamefont {Dumitrescu}, \citenamefont {Channa}, \citenamefont {Griffin},
  \citenamefont {Neaton}, \citenamefont {Potter},\ and\ \citenamefont
  {Analytis}}]{Nair2017}%
  \BibitemOpen
  \bibfield  {author} {\bibinfo {author} {\bibfnamefont {N.~L.}\ \bibnamefont
  {Nair}}, \bibinfo {author} {\bibfnamefont {P.~T.}\ \bibnamefont
  {Dumitrescu}}, \bibinfo {author} {\bibfnamefont {S.}~\bibnamefont {Channa}},
  \bibinfo {author} {\bibfnamefont {S.~M.}\ \bibnamefont {Griffin}}, \bibinfo
  {author} {\bibfnamefont {J.~B.}\ \bibnamefont {Neaton}}, \bibinfo {author}
  {\bibfnamefont {A.~C.}\ \bibnamefont {Potter}}, \ and\ \bibinfo {author}
  {\bibfnamefont {J.~G.}\ \bibnamefont {Analytis}},\ }\href {\doibase
  10.1103/PhysRevB.97.041111} {\bibfield  {journal} {\bibinfo  {journal} {Phys.
  Rev. B 97, 041111 (2018)}\ } (\bibinfo {year} {2017}),\
  10.1103/PhysRevB.97.041111},\ \Eprint {http://arxiv.org/abs/1708.03320}
  {arXiv:1708.03320 [cond-mat.mes-hall]} \BibitemShut {NoStop}%
\bibitem [{\citenamefont {Zhang}\ \emph {et~al.}(2017)\citenamefont {Zhang},
  \citenamefont {Wang}, \citenamefont {Yu}, \citenamefont {Liu}, \citenamefont
  {Liang}, \citenamefont {Huang}, \citenamefont {Nie}, \citenamefont {Sun},
  \citenamefont {Zhang}, \citenamefont {Shen}, \citenamefont {Liu},
  \citenamefont {Weng}, \citenamefont {Zhao}, \citenamefont {Chen},
  \citenamefont {Jia}, \citenamefont {Hu}, \citenamefont {Ding}, \citenamefont
  {Zhao}, \citenamefont {Gao}, \citenamefont {Li}, \citenamefont {He},
  \citenamefont {Zhao}, \citenamefont {Zhang}, \citenamefont {Zhang},
  \citenamefont {Yang}, \citenamefont {Wang}, \citenamefont {Peng},
  \citenamefont {Dai}, \citenamefont {Fang}, \citenamefont {Xu}, \citenamefont
  {Chen},\ and\ \citenamefont {Zhou}}]{Zhang2017}%
  \BibitemOpen
  \bibfield  {author} {\bibinfo {author} {\bibfnamefont {Y.}~\bibnamefont
  {Zhang}}, \bibinfo {author} {\bibfnamefont {C.}~\bibnamefont {Wang}},
  \bibinfo {author} {\bibfnamefont {L.}~\bibnamefont {Yu}}, \bibinfo {author}
  {\bibfnamefont {G.}~\bibnamefont {Liu}}, \bibinfo {author} {\bibfnamefont
  {A.}~\bibnamefont {Liang}}, \bibinfo {author} {\bibfnamefont
  {J.}~\bibnamefont {Huang}}, \bibinfo {author} {\bibfnamefont
  {S.}~\bibnamefont {Nie}}, \bibinfo {author} {\bibfnamefont {X.}~\bibnamefont
  {Sun}}, \bibinfo {author} {\bibfnamefont {Y.}~\bibnamefont {Zhang}}, \bibinfo
  {author} {\bibfnamefont {B.}~\bibnamefont {Shen}}, \bibinfo {author}
  {\bibfnamefont {J.}~\bibnamefont {Liu}}, \bibinfo {author} {\bibfnamefont
  {H.}~\bibnamefont {Weng}}, \bibinfo {author} {\bibfnamefont {L.}~\bibnamefont
  {Zhao}}, \bibinfo {author} {\bibfnamefont {G.}~\bibnamefont {Chen}}, \bibinfo
  {author} {\bibfnamefont {X.}~\bibnamefont {Jia}}, \bibinfo {author}
  {\bibfnamefont {C.}~\bibnamefont {Hu}}, \bibinfo {author} {\bibfnamefont
  {Y.}~\bibnamefont {Ding}}, \bibinfo {author} {\bibfnamefont {W.}~\bibnamefont
  {Zhao}}, \bibinfo {author} {\bibfnamefont {Q.}~\bibnamefont {Gao}}, \bibinfo
  {author} {\bibfnamefont {C.}~\bibnamefont {Li}}, \bibinfo {author}
  {\bibfnamefont {S.}~\bibnamefont {He}}, \bibinfo {author} {\bibfnamefont
  {L.}~\bibnamefont {Zhao}}, \bibinfo {author} {\bibfnamefont {F.}~\bibnamefont
  {Zhang}}, \bibinfo {author} {\bibfnamefont {S.}~\bibnamefont {Zhang}},
  \bibinfo {author} {\bibfnamefont {F.}~\bibnamefont {Yang}}, \bibinfo {author}
  {\bibfnamefont {Z.}~\bibnamefont {Wang}}, \bibinfo {author} {\bibfnamefont
  {Q.}~\bibnamefont {Peng}}, \bibinfo {author} {\bibfnamefont {X.}~\bibnamefont
  {Dai}}, \bibinfo {author} {\bibfnamefont {Z.}~\bibnamefont {Fang}}, \bibinfo
  {author} {\bibfnamefont {Z.}~\bibnamefont {Xu}}, \bibinfo {author}
  {\bibfnamefont {C.}~\bibnamefont {Chen}}, \ and\ \bibinfo {author}
  {\bibfnamefont {X.~J.}\ \bibnamefont {Zhou}},\ }\href {\doibase
  10.1038/ncomms15512} {\bibfield  {journal} {\bibinfo  {journal} {Nature
  Communications}\ }\textbf {\bibinfo {volume} {8}},\ \bibinfo {pages} {15512}
  (\bibinfo {year} {2017})}\BibitemShut {NoStop}%
\bibitem [{\citenamefont {Moreschini}\ \emph {et~al.}(2016)\citenamefont
  {Moreschini}, \citenamefont {Johannsen}, \citenamefont {Berger},
  \citenamefont {Denlinger}, \citenamefont {Jozwiak}, \citenamefont
  {Rotenberg}, \citenamefont {Kim}, \citenamefont {Bostwick},\ and\
  \citenamefont {Grioni}}]{Moreschini2016}%
  \BibitemOpen
  \bibfield  {author} {\bibinfo {author} {\bibfnamefont {L.}~\bibnamefont
  {Moreschini}}, \bibinfo {author} {\bibfnamefont {J.~C.}\ \bibnamefont
  {Johannsen}}, \bibinfo {author} {\bibfnamefont {H.}~\bibnamefont {Berger}},
  \bibinfo {author} {\bibfnamefont {J.}~\bibnamefont {Denlinger}}, \bibinfo
  {author} {\bibfnamefont {C.}~\bibnamefont {Jozwiak}}, \bibinfo {author}
  {\bibfnamefont {E.}~\bibnamefont {Rotenberg}}, \bibinfo {author}
  {\bibfnamefont {K.~S.}\ \bibnamefont {Kim}}, \bibinfo {author} {\bibfnamefont
  {A.}~\bibnamefont {Bostwick}}, \ and\ \bibinfo {author} {\bibfnamefont
  {M.}~\bibnamefont {Grioni}},\ }\href {\doibase 10.1103/PhysRevB.94.081101}
  {\bibfield  {journal} {\bibinfo  {journal} {Physical Review B}\ }\textbf
  {\bibinfo {volume} {94}},\ \bibinfo {pages} {081101} (\bibinfo {year}
  {2016})}\BibitemShut {NoStop}%
\bibitem [{\citenamefont {Xiong}\ \emph {et~al.}(2017)\citenamefont {Xiong},
  \citenamefont {Sobota}, \citenamefont {Yang}, \citenamefont {Soifer},
  \citenamefont {Gauthier}, \citenamefont {Lu}, \citenamefont {Lv},
  \citenamefont {Yao}, \citenamefont {Lu}, \citenamefont {Hashimoto},
  \citenamefont {Kirchmann}, \citenamefont {Chen},\ and\ \citenamefont
  {Shen}}]{Xiong2017}%
  \BibitemOpen
  \bibfield  {author} {\bibinfo {author} {\bibfnamefont {H.}~\bibnamefont
  {Xiong}}, \bibinfo {author} {\bibfnamefont {J.~A.}\ \bibnamefont {Sobota}},
  \bibinfo {author} {\bibfnamefont {S.-L.}\ \bibnamefont {Yang}}, \bibinfo
  {author} {\bibfnamefont {H.}~\bibnamefont {Soifer}}, \bibinfo {author}
  {\bibfnamefont {A.}~\bibnamefont {Gauthier}}, \bibinfo {author}
  {\bibfnamefont {M.-H.}\ \bibnamefont {Lu}}, \bibinfo {author} {\bibfnamefont
  {Y.-Y.}\ \bibnamefont {Lv}}, \bibinfo {author} {\bibfnamefont {S.-H.}\
  \bibnamefont {Yao}}, \bibinfo {author} {\bibfnamefont {D.}~\bibnamefont
  {Lu}}, \bibinfo {author} {\bibfnamefont {M.}~\bibnamefont {Hashimoto}},
  \bibinfo {author} {\bibfnamefont {P.~S.}\ \bibnamefont {Kirchmann}}, \bibinfo
  {author} {\bibfnamefont {Y.-F.}\ \bibnamefont {Chen}}, \ and\ \bibinfo
  {author} {\bibfnamefont {Z.-X.}\ \bibnamefont {Shen}},\ }\href {\doibase
  10.1103/PhysRevB.95.195119} {\bibfield  {journal} {\bibinfo  {journal}
  {Physical Review B}\ }\textbf {\bibinfo {volume} {95}},\ \bibinfo {pages}
  {195119} (\bibinfo {year} {2017})}\BibitemShut {NoStop}%
\bibitem [{\citenamefont {Liang}\ \emph {et~al.}(2019)\citenamefont {Liang},
  \citenamefont {Zhang}, \citenamefont {Zhang},\ and\ \citenamefont
  {Zheng}}]{Liang:2018bdb}%
  \BibitemOpen
  \bibfield  {author} {\bibinfo {author} {\bibfnamefont {Z.-L.}\ \bibnamefont
  {Liang}}, \bibinfo {author} {\bibfnamefont {L.}~\bibnamefont {Zhang}},
  \bibinfo {author} {\bibfnamefont {P.}~\bibnamefont {Zhang}}, \ and\ \bibinfo
  {author} {\bibfnamefont {F.}~\bibnamefont {Zheng}},\ }\href {\doibase
  10.1007/JHEP01(2019)149} {\bibfield  {journal} {\bibinfo  {journal} {JHEP}\
  }\textbf {\bibinfo {volume} {01}},\ \bibinfo {pages} {149} (\bibinfo {year}
  {2019})},\ \Eprint {http://arxiv.org/abs/1810.13394} {arXiv:1810.13394
  [cond-mat.mtrl-sci]} \BibitemShut {NoStop}%
\bibitem [{\citenamefont {Monkhorst}\ and\ \citenamefont
  {Pack}(1976)}]{Monkhorst:1976zz}%
  \BibitemOpen
  \bibfield  {author} {\bibinfo {author} {\bibfnamefont {H.~J.}\ \bibnamefont
  {Monkhorst}}\ and\ \bibinfo {author} {\bibfnamefont {J.~D.}\ \bibnamefont
  {Pack}},\ }\href {\doibase 10.1103/PhysRevB.13.5188} {\bibfield  {journal}
  {\bibinfo  {journal} {Phys. Rev. B}\ }\textbf {\bibinfo {volume} {13}},\
  \bibinfo {pages} {5188} (\bibinfo {year} {1976})}\BibitemShut {NoStop}%
\bibitem [{\citenamefont {Giannozzi}\ \emph
  {et~al.}(2009{\natexlab{a}})\citenamefont {Giannozzi}, \citenamefont
  {Baroni}, \citenamefont {Bonini}, \citenamefont {Calandra}, \citenamefont
  {Car}, \citenamefont {Cavazzoni}, \citenamefont {Ceresoli}, \citenamefont
  {Chiarotti}, \citenamefont {Cococcioni}, \citenamefont {Dabo}, \citenamefont
  {Corso}, \citenamefont {de~Gironcoli}, \citenamefont {Fabris}, \citenamefont
  {Fratesi}, \citenamefont {Gebauer}, \citenamefont {Gerstmann}, \citenamefont
  {Gougoussis}, \citenamefont {Kokalj}, \citenamefont {Lazzeri}, \citenamefont
  {Martin-Samos}, \citenamefont {Marzari}, \citenamefont {Mauri}, \citenamefont
  {Mazzarello}, \citenamefont {Paolini}, \citenamefont {Pasquarello},
  \citenamefont {Paulatto}, \citenamefont {Sbraccia}, \citenamefont {Scandolo},
  \citenamefont {Sclauzero}, \citenamefont {Seitsonen}, \citenamefont
  {Smogunov}, \citenamefont {Umari},\ and\ \citenamefont
  {Wentzcovitch}}]{Giannozzi2009}%
  \BibitemOpen
  \bibfield  {author} {\bibinfo {author} {\bibfnamefont {P.}~\bibnamefont
  {Giannozzi}}, \bibinfo {author} {\bibfnamefont {S.}~\bibnamefont {Baroni}},
  \bibinfo {author} {\bibfnamefont {N.}~\bibnamefont {Bonini}}, \bibinfo
  {author} {\bibfnamefont {M.}~\bibnamefont {Calandra}}, \bibinfo {author}
  {\bibfnamefont {R.}~\bibnamefont {Car}}, \bibinfo {author} {\bibfnamefont
  {C.}~\bibnamefont {Cavazzoni}}, \bibinfo {author} {\bibfnamefont
  {D.}~\bibnamefont {Ceresoli}}, \bibinfo {author} {\bibfnamefont {G.~L.}\
  \bibnamefont {Chiarotti}}, \bibinfo {author} {\bibfnamefont {M.}~\bibnamefont
  {Cococcioni}}, \bibinfo {author} {\bibfnamefont {I.}~\bibnamefont {Dabo}},
  \bibinfo {author} {\bibfnamefont {A.~D.}\ \bibnamefont {Corso}}, \bibinfo
  {author} {\bibfnamefont {S.}~\bibnamefont {de~Gironcoli}}, \bibinfo {author}
  {\bibfnamefont {S.}~\bibnamefont {Fabris}}, \bibinfo {author} {\bibfnamefont
  {G.}~\bibnamefont {Fratesi}}, \bibinfo {author} {\bibfnamefont
  {R.}~\bibnamefont {Gebauer}}, \bibinfo {author} {\bibfnamefont
  {U.}~\bibnamefont {Gerstmann}}, \bibinfo {author} {\bibfnamefont
  {C.}~\bibnamefont {Gougoussis}}, \bibinfo {author} {\bibfnamefont
  {A.}~\bibnamefont {Kokalj}}, \bibinfo {author} {\bibfnamefont
  {M.}~\bibnamefont {Lazzeri}}, \bibinfo {author} {\bibfnamefont
  {L.}~\bibnamefont {Martin-Samos}}, \bibinfo {author} {\bibfnamefont
  {N.}~\bibnamefont {Marzari}}, \bibinfo {author} {\bibfnamefont
  {F.}~\bibnamefont {Mauri}}, \bibinfo {author} {\bibfnamefont
  {R.}~\bibnamefont {Mazzarello}}, \bibinfo {author} {\bibfnamefont
  {S.}~\bibnamefont {Paolini}}, \bibinfo {author} {\bibfnamefont
  {A.}~\bibnamefont {Pasquarello}}, \bibinfo {author} {\bibfnamefont
  {L.}~\bibnamefont {Paulatto}}, \bibinfo {author} {\bibfnamefont
  {C.}~\bibnamefont {Sbraccia}}, \bibinfo {author} {\bibfnamefont
  {S.}~\bibnamefont {Scandolo}}, \bibinfo {author} {\bibfnamefont
  {G.}~\bibnamefont {Sclauzero}}, \bibinfo {author} {\bibfnamefont {A.~P.}\
  \bibnamefont {Seitsonen}}, \bibinfo {author} {\bibfnamefont {A.}~\bibnamefont
  {Smogunov}}, \bibinfo {author} {\bibfnamefont {P.}~\bibnamefont {Umari}}, \
  and\ \bibinfo {author} {\bibfnamefont {R.~M.}\ \bibnamefont {Wentzcovitch}},\
  }\href {\doibase 10.1088/0953-8984/21/39/395502} {\bibfield  {journal}
  {\bibinfo  {journal} {Journal of Physics: Condensed Matter}\ }\textbf
  {\bibinfo {volume} {21}},\ \bibinfo {pages} {395502} (\bibinfo {year}
  {2009}{\natexlab{a}})}\BibitemShut {NoStop}%
\bibitem [{\citenamefont {Giannozzi}\ \emph {et~al.}(2017)\citenamefont
  {Giannozzi}, \citenamefont {Andreussi}, \citenamefont {Brumme}, \citenamefont
  {Bunau}, \citenamefont {Nardelli}, \citenamefont {Calandra}, \citenamefont
  {Car}, \citenamefont {Cavazzoni}, \citenamefont {Ceresoli}, \citenamefont
  {Cococcioni}, \citenamefont {Colonna}, \citenamefont {Carnimeo},
  \citenamefont {Corso}, \citenamefont {de~Gironcoli}, \citenamefont {Delugas},
  \citenamefont {DiStasio~Jr.}, \citenamefont {Ferretti}, \citenamefont
  {Floris}, \citenamefont {Fratesi}, \citenamefont {Fugallo}, \citenamefont
  {Gebauer}, \citenamefont {Gerstmann}, \citenamefont {Giustino}, \citenamefont
  {Gorni}, \citenamefont {Jia}, \citenamefont {Kawamura}, \citenamefont {Ko},
  \citenamefont {Kokalj}, \citenamefont {K{\"u}{\c c}{\"u}kbenli},
  \citenamefont {Lazzeri}, \citenamefont {Marsili}, \citenamefont {Marzari},
  \citenamefont {Mauri}, \citenamefont {Nguyen}, \citenamefont {Nguyen},
  \citenamefont {Otero-de-la Roza}, \citenamefont {Paulatto}, \citenamefont
  {Ponc{\'e}}, \citenamefont {Rocca}, \citenamefont {Sabatini}, \citenamefont
  {Santra}, \citenamefont {Schlipf}, \citenamefont {Seitsonen}, \citenamefont
  {Smogunov}, \citenamefont {Timrov}, \citenamefont {Thonhauser}, \citenamefont
  {Umari}, \citenamefont {Vast}, \citenamefont {Wu},\ and\ \citenamefont
  {Baroni}}]{Giannozzi2017}%
  \BibitemOpen
  \bibfield  {author} {\bibinfo {author} {\bibfnamefont {P.}~\bibnamefont
  {Giannozzi}}, \bibinfo {author} {\bibfnamefont {O.}~\bibnamefont
  {Andreussi}}, \bibinfo {author} {\bibfnamefont {T.}~\bibnamefont {Brumme}},
  \bibinfo {author} {\bibfnamefont {O.}~\bibnamefont {Bunau}}, \bibinfo
  {author} {\bibfnamefont {M.~B.}\ \bibnamefont {Nardelli}}, \bibinfo {author}
  {\bibfnamefont {M.}~\bibnamefont {Calandra}}, \bibinfo {author}
  {\bibfnamefont {R.}~\bibnamefont {Car}}, \bibinfo {author} {\bibfnamefont
  {C.}~\bibnamefont {Cavazzoni}}, \bibinfo {author} {\bibfnamefont
  {D.}~\bibnamefont {Ceresoli}}, \bibinfo {author} {\bibfnamefont
  {M.}~\bibnamefont {Cococcioni}}, \bibinfo {author} {\bibfnamefont
  {N.}~\bibnamefont {Colonna}}, \bibinfo {author} {\bibfnamefont
  {I.}~\bibnamefont {Carnimeo}}, \bibinfo {author} {\bibfnamefont {A.~D.}\
  \bibnamefont {Corso}}, \bibinfo {author} {\bibfnamefont {S.}~\bibnamefont
  {de~Gironcoli}}, \bibinfo {author} {\bibfnamefont {P.}~\bibnamefont
  {Delugas}}, \bibinfo {author} {\bibfnamefont {R.~A.}\ \bibnamefont
  {DiStasio~Jr.}}, \bibinfo {author} {\bibfnamefont {A.}~\bibnamefont
  {Ferretti}}, \bibinfo {author} {\bibfnamefont {A.}~\bibnamefont {Floris}},
  \bibinfo {author} {\bibfnamefont {G.}~\bibnamefont {Fratesi}}, \bibinfo
  {author} {\bibfnamefont {G.}~\bibnamefont {Fugallo}}, \bibinfo {author}
  {\bibfnamefont {R.}~\bibnamefont {Gebauer}}, \bibinfo {author} {\bibfnamefont
  {U.}~\bibnamefont {Gerstmann}}, \bibinfo {author} {\bibfnamefont
  {F.}~\bibnamefont {Giustino}}, \bibinfo {author} {\bibfnamefont
  {T.}~\bibnamefont {Gorni}}, \bibinfo {author} {\bibfnamefont
  {J.}~\bibnamefont {Jia}}, \bibinfo {author} {\bibfnamefont {M.}~\bibnamefont
  {Kawamura}}, \bibinfo {author} {\bibfnamefont {H.-Y.}\ \bibnamefont {Ko}},
  \bibinfo {author} {\bibfnamefont {A.}~\bibnamefont {Kokalj}}, \bibinfo
  {author} {\bibfnamefont {E.}~\bibnamefont {K{\"u}{\c c}{\"u}kbenli}},
  \bibinfo {author} {\bibfnamefont {M.}~\bibnamefont {Lazzeri}}, \bibinfo
  {author} {\bibfnamefont {M.}~\bibnamefont {Marsili}}, \bibinfo {author}
  {\bibfnamefont {N.}~\bibnamefont {Marzari}}, \bibinfo {author} {\bibfnamefont
  {F.}~\bibnamefont {Mauri}}, \bibinfo {author} {\bibfnamefont {N.~L.}\
  \bibnamefont {Nguyen}}, \bibinfo {author} {\bibfnamefont {H.-V.}\
  \bibnamefont {Nguyen}}, \bibinfo {author} {\bibfnamefont {A.}~\bibnamefont
  {Otero-de-la Roza}}, \bibinfo {author} {\bibfnamefont {L.}~\bibnamefont
  {Paulatto}}, \bibinfo {author} {\bibfnamefont {S.}~\bibnamefont {Ponc{\'e}}},
  \bibinfo {author} {\bibfnamefont {D.}~\bibnamefont {Rocca}}, \bibinfo
  {author} {\bibfnamefont {R.}~\bibnamefont {Sabatini}}, \bibinfo {author}
  {\bibfnamefont {B.}~\bibnamefont {Santra}}, \bibinfo {author} {\bibfnamefont
  {M.}~\bibnamefont {Schlipf}}, \bibinfo {author} {\bibfnamefont {A.~P.}\
  \bibnamefont {Seitsonen}}, \bibinfo {author} {\bibfnamefont {A.}~\bibnamefont
  {Smogunov}}, \bibinfo {author} {\bibfnamefont {I.}~\bibnamefont {Timrov}},
  \bibinfo {author} {\bibfnamefont {T.}~\bibnamefont {Thonhauser}}, \bibinfo
  {author} {\bibfnamefont {P.}~\bibnamefont {Umari}}, \bibinfo {author}
  {\bibfnamefont {N.}~\bibnamefont {Vast}}, \bibinfo {author} {\bibfnamefont
  {X.}~\bibnamefont {Wu}}, \ and\ \bibinfo {author} {\bibfnamefont
  {S.}~\bibnamefont {Baroni}},\ }\href {\doibase 10.1088/1361-648X/aa8f79}
  {\bibfield  {journal} {\bibinfo  {journal} {Journal of Physics: Condensed
  Matter}\ }\textbf {\bibinfo {volume} {29}},\ \bibinfo {pages} {465901}
  (\bibinfo {year} {2017})},\ \bibinfo {note} {arXiv: 1709.10010}\BibitemShut
  {NoStop}%
\bibitem [{\citenamefont {Giannozzi}\ \emph {et~al.}(2020)\citenamefont
  {Giannozzi}, \citenamefont {Baseggio}, \citenamefont {Bonf{\`{a}}},
  \citenamefont {Brunato}, \citenamefont {Car}, \citenamefont {Carnimeo},
  \citenamefont {Cavazzoni}, \citenamefont {de~Gironcoli}, \citenamefont
  {Delugas}, \citenamefont {Ruffino}, \citenamefont {Ferretti}, \citenamefont
  {Marzari}, \citenamefont {Timrov}, \citenamefont {Urru},\ and\ \citenamefont
  {Baroni}}]{Giannozzi2020}%
  \BibitemOpen
  \bibfield  {author} {\bibinfo {author} {\bibfnamefont {P.}~\bibnamefont
  {Giannozzi}}, \bibinfo {author} {\bibfnamefont {O.}~\bibnamefont {Baseggio}},
  \bibinfo {author} {\bibfnamefont {P.}~\bibnamefont {Bonf{\`{a}}}}, \bibinfo
  {author} {\bibfnamefont {D.}~\bibnamefont {Brunato}}, \bibinfo {author}
  {\bibfnamefont {R.}~\bibnamefont {Car}}, \bibinfo {author} {\bibfnamefont
  {I.}~\bibnamefont {Carnimeo}}, \bibinfo {author} {\bibfnamefont
  {C.}~\bibnamefont {Cavazzoni}}, \bibinfo {author} {\bibfnamefont
  {S.}~\bibnamefont {de~Gironcoli}}, \bibinfo {author} {\bibfnamefont
  {P.}~\bibnamefont {Delugas}}, \bibinfo {author} {\bibfnamefont {F.~F.}\
  \bibnamefont {Ruffino}}, \bibinfo {author} {\bibfnamefont {A.}~\bibnamefont
  {Ferretti}}, \bibinfo {author} {\bibfnamefont {N.}~\bibnamefont {Marzari}},
  \bibinfo {author} {\bibfnamefont {I.}~\bibnamefont {Timrov}}, \bibinfo
  {author} {\bibfnamefont {A.}~\bibnamefont {Urru}}, \ and\ \bibinfo {author}
  {\bibfnamefont {S.}~\bibnamefont {Baroni}},\ }\href {\doibase
  10.1063/5.0005082} {\bibfield  {journal} {\bibinfo  {journal} {The Journal of
  Chemical Physics}\ }\textbf {\bibinfo {volume} {152}},\ \bibinfo {pages}
  {154105} (\bibinfo {year} {2020})}\BibitemShut {NoStop}%
\bibitem [{\citenamefont {Griffin}\ \emph {et~al.}(2018)\citenamefont
  {Griffin}, \citenamefont {Knapen}, \citenamefont {Lin},\ and\ \citenamefont
  {Zurek}}]{Griffin:2018bjn}%
  \BibitemOpen
  \bibfield  {author} {\bibinfo {author} {\bibfnamefont {S.}~\bibnamefont
  {Griffin}}, \bibinfo {author} {\bibfnamefont {S.}~\bibnamefont {Knapen}},
  \bibinfo {author} {\bibfnamefont {T.}~\bibnamefont {Lin}}, \ and\ \bibinfo
  {author} {\bibfnamefont {K.~M.}\ \bibnamefont {Zurek}},\ }\href {\doibase
  10.1103/PhysRevD.98.115034} {\bibfield  {journal} {\bibinfo  {journal} {Phys.
  Rev. D}\ }\textbf {\bibinfo {volume} {98}},\ \bibinfo {pages} {115034}
  (\bibinfo {year} {2018})},\ \Eprint {http://arxiv.org/abs/1807.10291}
  {arXiv:1807.10291 [hep-ph]} \BibitemShut {NoStop}%
\bibitem [{\citenamefont {Knapen}\ \emph {et~al.}(2022)\citenamefont {Knapen},
  \citenamefont {Kozaczuk},\ and\ \citenamefont {Lin}}]{Knapen:2021bwg}%
  \BibitemOpen
  \bibfield  {author} {\bibinfo {author} {\bibfnamefont {S.}~\bibnamefont
  {Knapen}}, \bibinfo {author} {\bibfnamefont {J.}~\bibnamefont {Kozaczuk}}, \
  and\ \bibinfo {author} {\bibfnamefont {T.}~\bibnamefont {Lin}},\ }\href
  {\doibase 10.1103/PhysRevD.105.015014} {\bibfield  {journal} {\bibinfo
  {journal} {Phys. Rev. D}\ }\textbf {\bibinfo {volume} {105}},\ \bibinfo
  {pages} {015014} (\bibinfo {year} {2022})},\ \Eprint
  {http://arxiv.org/abs/2104.12786} {arXiv:2104.12786 [hep-ph]} \BibitemShut
  {NoStop}%
\bibitem [{\citenamefont {Adelberger}\ \emph {et~al.}(2003)\citenamefont
  {Adelberger}, \citenamefont {Heckel},\ and\ \citenamefont
  {Nelson}}]{Adelberger:2003zx}%
  \BibitemOpen
  \bibfield  {author} {\bibinfo {author} {\bibfnamefont {E.~G.}\ \bibnamefont
  {Adelberger}}, \bibinfo {author} {\bibfnamefont {B.~R.}\ \bibnamefont
  {Heckel}}, \ and\ \bibinfo {author} {\bibfnamefont {A.~E.}\ \bibnamefont
  {Nelson}},\ }\href {\doibase 10.1146/annurev.nucl.53.041002.110503}
  {\bibfield  {journal} {\bibinfo  {journal} {Ann. Rev. Nucl. Part. Sci.}\
  }\textbf {\bibinfo {volume} {53}},\ \bibinfo {pages} {77} (\bibinfo {year}
  {2003})},\ \Eprint {http://arxiv.org/abs/hep-ph/0307284}
  {arXiv:hep-ph/0307284} \BibitemShut {NoStop}%
\bibitem [{\citenamefont {Hardy}\ and\ \citenamefont
  {Lasenby}(2017)}]{Hardy:2016kme}%
  \BibitemOpen
  \bibfield  {author} {\bibinfo {author} {\bibfnamefont {E.}~\bibnamefont
  {Hardy}}\ and\ \bibinfo {author} {\bibfnamefont {R.}~\bibnamefont
  {Lasenby}},\ }\href {\doibase 10.1007/JHEP02(2017)033} {\bibfield  {journal}
  {\bibinfo  {journal} {JHEP}\ }\textbf {\bibinfo {volume} {02}},\ \bibinfo
  {pages} {033} (\bibinfo {year} {2017})},\ \Eprint
  {http://arxiv.org/abs/1611.05852} {arXiv:1611.05852 [hep-ph]} \BibitemShut
  {NoStop}%
\bibitem [{\citenamefont {Miller~Bertolami}\ \emph {et~al.}(2014)\citenamefont
  {Miller~Bertolami}, \citenamefont {Melendez}, \citenamefont {Althaus},\ and\
  \citenamefont {Isern}}]{MillerBertolami:2014rka}%
  \BibitemOpen
  \bibfield  {author} {\bibinfo {author} {\bibfnamefont {M.~M.}\ \bibnamefont
  {Miller~Bertolami}}, \bibinfo {author} {\bibfnamefont {B.~E.}\ \bibnamefont
  {Melendez}}, \bibinfo {author} {\bibfnamefont {L.~G.}\ \bibnamefont
  {Althaus}}, \ and\ \bibinfo {author} {\bibfnamefont {J.}~\bibnamefont
  {Isern}},\ }\href {\doibase 10.1088/1475-7516/2014/10/069} {\bibfield
  {journal} {\bibinfo  {journal} {JCAP}\ }\textbf {\bibinfo {volume} {10}},\
  \bibinfo {pages} {069} (\bibinfo {year} {2014})},\ \Eprint
  {http://arxiv.org/abs/1406.7712} {arXiv:1406.7712 [hep-ph]} \BibitemShut
  {NoStop}%
\bibitem [{\citenamefont {Tanabashi}\ \emph {et~al.}(2018)\citenamefont
  {Tanabashi} \emph {et~al.}}]{ParticleDataGroup:2018ovx}%
  \BibitemOpen
  \bibfield  {author} {\bibinfo {author} {\bibfnamefont {M.}~\bibnamefont
  {Tanabashi}} \emph {et~al.} (\bibinfo {collaboration} {Particle Data
  Group}),\ }\href {\doibase 10.1103/PhysRevD.98.030001} {\bibfield  {journal}
  {\bibinfo  {journal} {Phys. Rev. D}\ }\textbf {\bibinfo {volume} {98}},\
  \bibinfo {pages} {030001} (\bibinfo {year} {2018})}\BibitemShut {NoStop}%
\bibitem [{\citenamefont {Trickle}\ \emph {et~al.}(2022)\citenamefont
  {Trickle}, \citenamefont {Zhang},\ and\ \citenamefont
  {Zurek}}]{Trickle:2020oki}%
  \BibitemOpen
  \bibfield  {author} {\bibinfo {author} {\bibfnamefont {T.}~\bibnamefont
  {Trickle}}, \bibinfo {author} {\bibfnamefont {Z.}~\bibnamefont {Zhang}}, \
  and\ \bibinfo {author} {\bibfnamefont {K.~M.}\ \bibnamefont {Zurek}},\ }\href
  {\doibase 10.1103/PhysRevD.105.015001} {\bibfield  {journal} {\bibinfo
  {journal} {Phys. Rev. D}\ }\textbf {\bibinfo {volume} {105}},\ \bibinfo
  {pages} {015001} (\bibinfo {year} {2022})},\ \Eprint
  {http://arxiv.org/abs/2009.13534} {arXiv:2009.13534 [hep-ph]} \BibitemShut
  {NoStop}%
\bibitem [{\citenamefont {Hochberg}\ \emph {et~al.}(2021)\citenamefont
  {Hochberg}, \citenamefont {Kahn}, \citenamefont {Kurinsky}, \citenamefont
  {Lehmann}, \citenamefont {Yu},\ and\ \citenamefont
  {Berggren}}]{Hochberg2021}%
  \BibitemOpen
  \bibfield  {author} {\bibinfo {author} {\bibfnamefont {Y.}~\bibnamefont
  {Hochberg}}, \bibinfo {author} {\bibfnamefont {Y.}~\bibnamefont {Kahn}},
  \bibinfo {author} {\bibfnamefont {N.}~\bibnamefont {Kurinsky}}, \bibinfo
  {author} {\bibfnamefont {B.~V.}\ \bibnamefont {Lehmann}}, \bibinfo {author}
  {\bibfnamefont {T.~C.}\ \bibnamefont {Yu}}, \ and\ \bibinfo {author}
  {\bibfnamefont {K.~K.}\ \bibnamefont {Berggren}},\ }\href {\doibase
  10.1103/PhysRevLett.127.151802} {\bibfield  {journal} {\bibinfo  {journal}
  {Phys. Rev. Lett.}\ }\textbf {\bibinfo {volume} {127}},\ \bibinfo {pages}
  {151802} (\bibinfo {year} {2021})},\ \Eprint
  {http://arxiv.org/abs/2101.08263} {arXiv:2101.08263 [hep-ph]} \BibitemShut
  {NoStop}%
\bibitem [{\citenamefont {Vogel}\ and\ \citenamefont
  {Redondo}(2014)}]{Vogel:2013raa}%
  \BibitemOpen
  \bibfield  {author} {\bibinfo {author} {\bibfnamefont {H.}~\bibnamefont
  {Vogel}}\ and\ \bibinfo {author} {\bibfnamefont {J.}~\bibnamefont
  {Redondo}},\ }\href {\doibase 10.1088/1475-7516/2014/02/029} {\bibfield
  {journal} {\bibinfo  {journal} {JCAP}\ }\textbf {\bibinfo {volume} {02}},\
  \bibinfo {pages} {029} (\bibinfo {year} {2014})},\ \Eprint
  {http://arxiv.org/abs/1311.2600} {arXiv:1311.2600 [hep-ph]} \BibitemShut
  {NoStop}%
\bibitem [{\citenamefont {Dvorkin}\ \emph {et~al.}(2019)\citenamefont
  {Dvorkin}, \citenamefont {Lin},\ and\ \citenamefont
  {Schutz}}]{Dvorkin:2019zdi}%
  \BibitemOpen
  \bibfield  {author} {\bibinfo {author} {\bibfnamefont {C.}~\bibnamefont
  {Dvorkin}}, \bibinfo {author} {\bibfnamefont {T.}~\bibnamefont {Lin}}, \ and\
  \bibinfo {author} {\bibfnamefont {K.}~\bibnamefont {Schutz}},\ }\href
  {\doibase 10.1103/PhysRevD.99.115009} {\bibfield  {journal} {\bibinfo
  {journal} {Phys. Rev. D}\ }\textbf {\bibinfo {volume} {99}},\ \bibinfo
  {pages} {115009} (\bibinfo {year} {2019})},\ \Eprint
  {http://arxiv.org/abs/1902.08623} {arXiv:1902.08623 [hep-ph]} \BibitemShut
  {NoStop}%
\bibitem [{\citenamefont {Perdew}\ \emph {et~al.}(1996)\citenamefont {Perdew},
  \citenamefont {Burke},\ and\ \citenamefont
  {Ernzerhof}}]{perdew1996generalized}%
  \BibitemOpen
  \bibfield  {author} {\bibinfo {author} {\bibfnamefont {J.~P.}\ \bibnamefont
  {Perdew}}, \bibinfo {author} {\bibfnamefont {K.}~\bibnamefont {Burke}}, \
  and\ \bibinfo {author} {\bibfnamefont {M.}~\bibnamefont {Ernzerhof}},\ }\href
  {\doibase 10.1103/PhysRevLett.77.3865} {\bibfield  {journal} {\bibinfo
  {journal} {Phys. Rev. Lett.}\ }\textbf {\bibinfo {volume} {77}},\ \bibinfo
  {pages} {3865} (\bibinfo {year} {1996})}\BibitemShut {NoStop}%
\bibitem [{\citenamefont {Giannozzi}\ \emph
  {et~al.}(2009{\natexlab{b}})\citenamefont {Giannozzi}, \citenamefont
  {Baroni}, \citenamefont {Bonini}, \citenamefont {Calandra}, \citenamefont
  {Car}, \citenamefont {Cavazzoni}, \citenamefont {Ceresoli}, \citenamefont
  {Chiarotti}, \citenamefont {Cococcioni}, \citenamefont {Dabo} \emph
  {et~al.}}]{giannozzi2009quantum}%
  \BibitemOpen
  \bibfield  {author} {\bibinfo {author} {\bibfnamefont {P.}~\bibnamefont
  {Giannozzi}}, \bibinfo {author} {\bibfnamefont {S.}~\bibnamefont {Baroni}},
  \bibinfo {author} {\bibfnamefont {N.}~\bibnamefont {Bonini}}, \bibinfo
  {author} {\bibfnamefont {M.}~\bibnamefont {Calandra}}, \bibinfo {author}
  {\bibfnamefont {R.}~\bibnamefont {Car}}, \bibinfo {author} {\bibfnamefont
  {C.}~\bibnamefont {Cavazzoni}}, \bibinfo {author} {\bibfnamefont
  {D.}~\bibnamefont {Ceresoli}}, \bibinfo {author} {\bibfnamefont {G.~L.}\
  \bibnamefont {Chiarotti}}, \bibinfo {author} {\bibfnamefont {M.}~\bibnamefont
  {Cococcioni}}, \bibinfo {author} {\bibfnamefont {I.}~\bibnamefont {Dabo}},
  \emph {et~al.},\ }\href@noop {} {\bibfield  {journal} {\bibinfo  {journal}
  {Journal of physics: Condensed matter}\ }\textbf {\bibinfo {volume} {21}},\
  \bibinfo {pages} {395502} (\bibinfo {year} {2009}{\natexlab{b}})}\BibitemShut
  {NoStop}%
\bibitem [{\citenamefont {Fjellv{\aa}g}\ and\ \citenamefont
  {Kjekshus}(1986)}]{fjellvaag1986structural}%
  \BibitemOpen
  \bibfield  {author} {\bibinfo {author} {\bibfnamefont {H.}~\bibnamefont
  {Fjellv{\aa}g}}\ and\ \bibinfo {author} {\bibfnamefont {A.}~\bibnamefont
  {Kjekshus}},\ }\href@noop {} {\bibfield  {journal} {\bibinfo  {journal}
  {Solid State Communications}\ }\textbf {\bibinfo {volume} {60}},\ \bibinfo
  {pages} {91} (\bibinfo {year} {1986})}\BibitemShut {NoStop}%
\bibitem [{\citenamefont {Perdew}\ and\ \citenamefont
  {Zunger}(1981)}]{perdew1981self}%
  \BibitemOpen
  \bibfield  {author} {\bibinfo {author} {\bibfnamefont {J.~P.}\ \bibnamefont
  {Perdew}}\ and\ \bibinfo {author} {\bibfnamefont {A.}~\bibnamefont
  {Zunger}},\ }\href {\doibase 10.1103/PhysRevB.23.5048} {\bibfield  {journal}
  {\bibinfo  {journal} {Phys. Rev. B}\ }\textbf {\bibinfo {volume} {23}},\
  \bibinfo {pages} {5048} (\bibinfo {year} {1981})}\BibitemShut {NoStop}%
\bibitem [{\citenamefont {Troullier}\ and\ \citenamefont
  {Martins}(1991)}]{troullier1991efficient}%
  \BibitemOpen
  \bibfield  {author} {\bibinfo {author} {\bibfnamefont {N.}~\bibnamefont
  {Troullier}}\ and\ \bibinfo {author} {\bibfnamefont {J.~L.}\ \bibnamefont
  {Martins}},\ }\href {\doibase 10.1103/PhysRevB.43.1993} {\bibfield  {journal}
  {\bibinfo  {journal} {Phys. Rev. B}\ }\textbf {\bibinfo {volume} {43}},\
  \bibinfo {pages} {1993} (\bibinfo {year} {1991})}\BibitemShut {NoStop}%
\bibitem [{\citenamefont {van Setten}\ \emph {et~al.}(2018)\citenamefont {van
  Setten}, \citenamefont {Giantomassi}, \citenamefont {Bousquet}, \citenamefont
  {Verstraete}, \citenamefont {Hamann}, \citenamefont {Gonze},\ and\
  \citenamefont {Rignanese}}]{van2018pseudodojo}%
  \BibitemOpen
  \bibfield  {author} {\bibinfo {author} {\bibfnamefont {M.~J.}\ \bibnamefont
  {van Setten}}, \bibinfo {author} {\bibfnamefont {M.}~\bibnamefont
  {Giantomassi}}, \bibinfo {author} {\bibfnamefont {E.}~\bibnamefont
  {Bousquet}}, \bibinfo {author} {\bibfnamefont {M.~J.}\ \bibnamefont
  {Verstraete}}, \bibinfo {author} {\bibfnamefont {D.~R.}\ \bibnamefont
  {Hamann}}, \bibinfo {author} {\bibfnamefont {X.}~\bibnamefont {Gonze}}, \
  and\ \bibinfo {author} {\bibfnamefont {G.-M.}\ \bibnamefont {Rignanese}},\
  }\href@noop {} {\bibfield  {journal} {\bibinfo  {journal} {Computer Physics
  Communications}\ }\textbf {\bibinfo {volume} {226}},\ \bibinfo {pages} {39}
  (\bibinfo {year} {2018})}\BibitemShut {NoStop}%
\bibitem [{\citenamefont {Fan}\ \emph {et~al.}(2017)\citenamefont {Fan},
  \citenamefont {Liang}, \citenamefont {Chen}, \citenamefont {Yao},\ and\
  \citenamefont {Zhou}}]{fan2017transition}%
  \BibitemOpen
  \bibfield  {author} {\bibinfo {author} {\bibfnamefont {Z.}~\bibnamefont
  {Fan}}, \bibinfo {author} {\bibfnamefont {Q.-F.}\ \bibnamefont {Liang}},
  \bibinfo {author} {\bibfnamefont {Y.}~\bibnamefont {Chen}}, \bibinfo {author}
  {\bibfnamefont {S.-H.}\ \bibnamefont {Yao}}, \ and\ \bibinfo {author}
  {\bibfnamefont {J.}~\bibnamefont {Zhou}},\ }\href@noop {} {\bibfield
  {journal} {\bibinfo  {journal} {Scientific reports}\ }\textbf {\bibinfo
  {volume} {7}},\ \bibinfo {pages} {1} (\bibinfo {year} {2017})}\BibitemShut
  {NoStop}%
\bibitem [{\citenamefont {Sangalli}\ \emph {et~al.}(2017)\citenamefont
  {Sangalli}, \citenamefont {Berger}, \citenamefont {Attaccalite},
  \citenamefont {Gr\"uning},\ and\ \citenamefont
  {Romaniello}}]{PhysRevB.95.155203}%
  \BibitemOpen
  \bibfield  {author} {\bibinfo {author} {\bibfnamefont {D.}~\bibnamefont
  {Sangalli}}, \bibinfo {author} {\bibfnamefont {J.~A.}\ \bibnamefont
  {Berger}}, \bibinfo {author} {\bibfnamefont {C.}~\bibnamefont {Attaccalite}},
  \bibinfo {author} {\bibfnamefont {M.}~\bibnamefont {Gr\"uning}}, \ and\
  \bibinfo {author} {\bibfnamefont {P.}~\bibnamefont {Romaniello}},\ }\href
  {\doibase 10.1103/PhysRevB.95.155203} {\bibfield  {journal} {\bibinfo
  {journal} {Phys. Rev. B}\ }\textbf {\bibinfo {volume} {95}},\ \bibinfo
  {pages} {155203} (\bibinfo {year} {2017})}\BibitemShut {NoStop}%
\bibitem [{\citenamefont {Lozovik}\ \emph {et~al.}(2008)\citenamefont
  {Lozovik}, \citenamefont {Merkulova}, \citenamefont {Sokolik}, \citenamefont
  {Morozov}, \citenamefont {Novoselov},\ and\ \citenamefont
  {Geim}}]{Lozovik2008}%
  \BibitemOpen
  \bibfield  {author} {\bibinfo {author} {\bibfnamefont {Y.~E.}\ \bibnamefont
  {Lozovik}}, \bibinfo {author} {\bibfnamefont {S.~P.}\ \bibnamefont
  {Merkulova}}, \bibinfo {author} {\bibfnamefont {A.~A.}\ \bibnamefont
  {Sokolik}}, \bibinfo {author} {\bibfnamefont {S.~V.}\ \bibnamefont
  {Morozov}}, \bibinfo {author} {\bibfnamefont {K.~S.}\ \bibnamefont
  {Novoselov}}, \ and\ \bibinfo {author} {\bibfnamefont {A.~K.}\ \bibnamefont
  {Geim}},\ }\href {\doibase 10.1070/pu2008v051n07abeh006793} {\bibfield
  {journal} {\bibinfo  {journal} {Physics-Uspekhi}\ }\textbf {\bibinfo {volume}
  {51}},\ \bibinfo {pages} {727} (\bibinfo {year} {2008})}\BibitemShut
  {NoStop}%
\end{thebibliography}%

\end{document}